\documentclass[twocolumn,nofootinbib,jcp,preprintnumbers,amsmath,amssymb]{revtex4-1}
\usepackage{natbib,hyperref}
\usepackage{amsmath}
\usepackage{graphicx}
\usepackage{hyperref} 
\usepackage{dcolumn}
\usepackage{bm}
\usepackage{epsfig,color,xspace,multirow,xr,bbold}
\usepackage[all]{xy}
\usepackage{setspace}
\usepackage{url}
\usepackage{soul}
\usepackage[colorinlistoftodos]{todonotes}
\usepackage{threeparttable} 
\usepackage{natbib,hyperref}
\usepackage{float}
\usepackage{placeins}
\usepackage[utf8]{inputenc}

\usepackage{xcolor}

\usepackage{soul} % for highlighting
\begin{document}
%\preprint{AIP/123-QED}
\title{
Resilience of Hund's rule in the Chemical Space of Small Organic Molecules
}
\date{\today}

\author{Atreyee Majumdar}
\author{Raghunathan Ramakrishnan}
\email{ramakrishnan@tifrh.res.in}
\affiliation{$^1$Tata Institute of Fundamental Research, Hyderabad 500046, India}

\keywords{
S1-T1 energy gap,
Hund's rule,
double-hybrid DFT,
ADC($n$),
basis set,
exchange-correlation
}

\begin{abstract}
\noindent
We embark on a quest to identify small molecules in the chemical space that can potentially violate Hund's rule. 
Utilizing twelve TDDFT approximations and the ADC(2) many-body method, we report the energies of S$_1$ and T$_1$ excited states of 12,880 closed-shell organic molecules within the bigQM7$\omega$ dataset with up to 7 CONF atoms.
In this comprehensive dataset, none of the molecules, in their minimum energy geometry, exhibit a negative S$_1$-T$_1$ energy gap at the ADC($2$) level while several molecules display values $<0.1$ eV. 
The spin-component-scaled double-hybrid method, SCS-PBE-QIDH, 
demonstrates the best agreement with ADC(2). 
Yet, at this level, a few molecules with a strained $sp^3$-N center turn out as false-positives with the S$_1$ state lower in energy than T$_1$. 
We investigate a prototypical cage molecule 
with an energy gap $<-0.2$ eV, which a closer examination revealed as another false positive. 
We conclude that in the chemical space of small closed-shell organic molecules, it is possible to identify geometric and electronic structural features giving rise to S$_1$-T$_1$ degeneracy; still, there is no evidence of a negative gap. 
We share the dataset generated for this study as a module, 
to facilitate seamless molecular discovery through data mining.
\end{abstract}

\maketitle
\section{Introduction}
Studies from the 1970s onward, some even titled ``Violation of Hund's (multiplicity) rule$\ldots$'' have explored the possibility of an electronic state of lower spin-multiplicity being more stable than its higher multiplicity counterpart \cite{kollmar1978violation, koseki1985violation, borden1994violations, toyota1986violation, toyota1988violation, hrovat1997violations, sancho2022violation}. Although no experimentally known cases of Hund's rule violation exist among molecules in their ground state, 
{\it ab initio} wavefunction models have suggested that dynamic spin polarization effects may favor open-shell singlets over triplets\cite{kollmar1978violation}. 
In 1980, Leupin {\it et al.} suggested the likelihood of a more stable lowest excited singlet state, S$_1$, 
compared to the triplet state, T$_1$, based on fluorescence measurements of cycl[3.3.3]azines\cite{leupin1980low}. 
Likewise, in 1985, certain non-alternant polycyclic hydrocarbons 
were considered to exhibit negative S$_1$-T$_1$ gaps (STGs)\cite{koseki1985violation}. Typically, S$_1$ and T$_1$ are represented primarily by singly excited configurations, $^1 \chi_{a \rightarrow r}$ and $^3 \chi_{a \rightarrow r}$, where $a$ and $r$ are the occupied and virtual molecular orbitals (MOs) determined using the self-consistent-field (SCF) method for the ground state configuration, $^1 \chi_0$. Hence, the expressions for the excitation energies are $E(^1 \chi_{a \rightarrow r})-E(^1 \chi_0)=\varepsilon_r-\varepsilon_a-J_{ar}+2K_{ar}$ and $E(^3 \chi_{a \rightarrow r})-E(^1 \chi_0)=\varepsilon_r-\varepsilon_a-J_{ar}$, indicating 
${\rm STG}=E(^1 \chi_{a \rightarrow r})-E(^3 \chi_{a \rightarrow r})=2K_{ar}$ (twice the exchange integral). As the overlap between the densities of the $a$ and $r$ MOs diminishes, a reduction in $K_{ar}$ leads to degenerate S$_1$ and T$_1$\cite{leupin1980low}. Along with vanishing $K_{ar}$,
a decrease in the percentage contribution of the $^1 \chi_{a \rightarrow r}$ configuration and 
an increased contribution from the doubly-excited configuration,
$^1 \chi_{aa \rightarrow rr}$, was hypothesized to be a factor to
selectively stabilize the S$_1$ state over T$_1$ resulting in a negative ${\rm STG}$\cite{bonacic1985charge}.

\begin{figure}[hb]
\centering
\includegraphics[width=\linewidth]{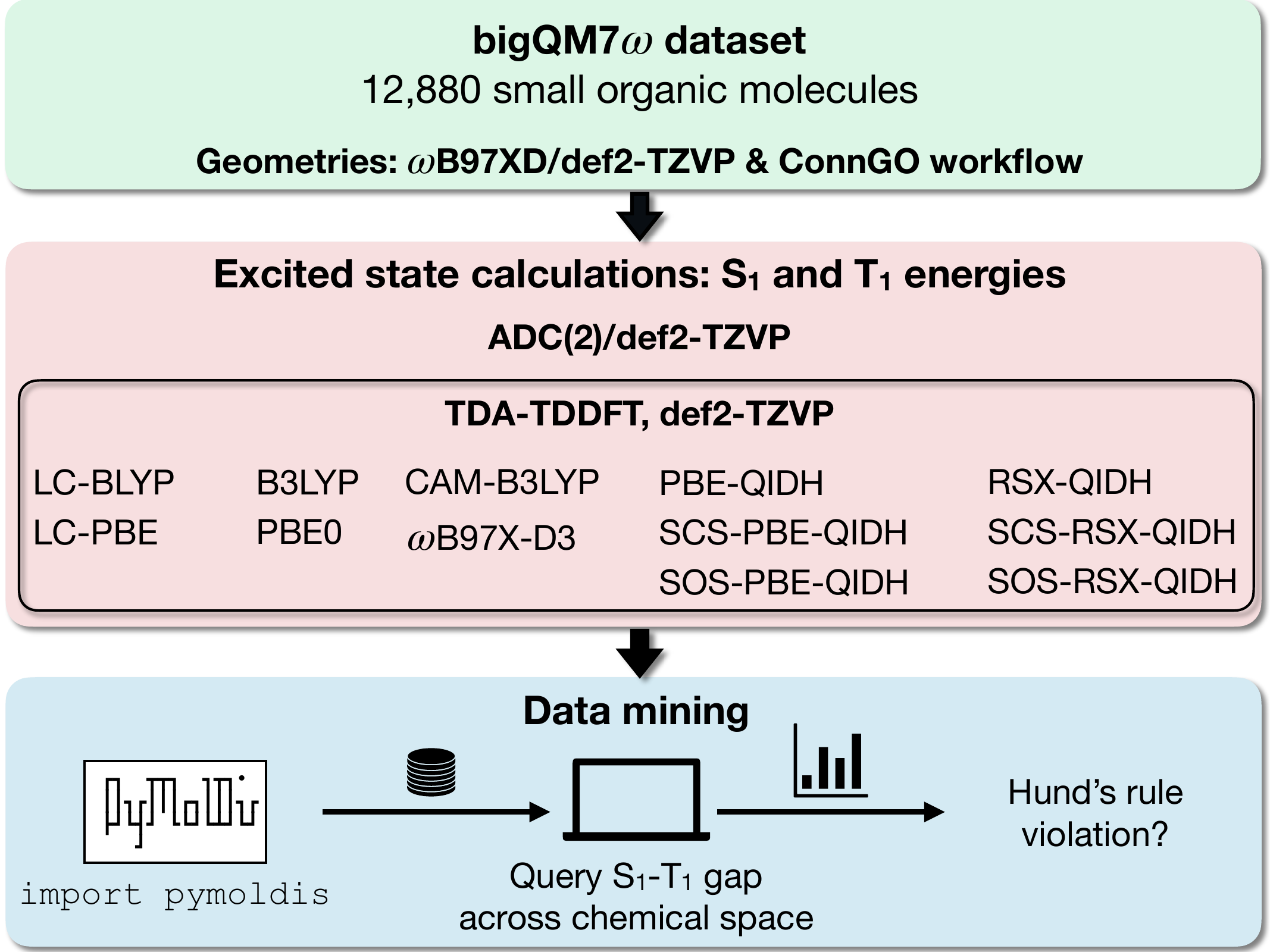}
\caption{Workflow outlining the design of the {\tt pymoldis} module\cite{pymoldis} for data mining 
S$_1$ and T$_1$ energies across 12,880
molecules with up to 7 CONF atoms in the bigQM7$\omega$ dataset\cite{kayastha2022resolution}. ADC(2) and TDDFT calculations were performed as a part of the present work.
See Supplementary Information (SI) for screenshots of example queries.
}
\label{fig:workflow}
\end{figure}
\begin{figure*}[ht]
\centering
\includegraphics[width=\linewidth]{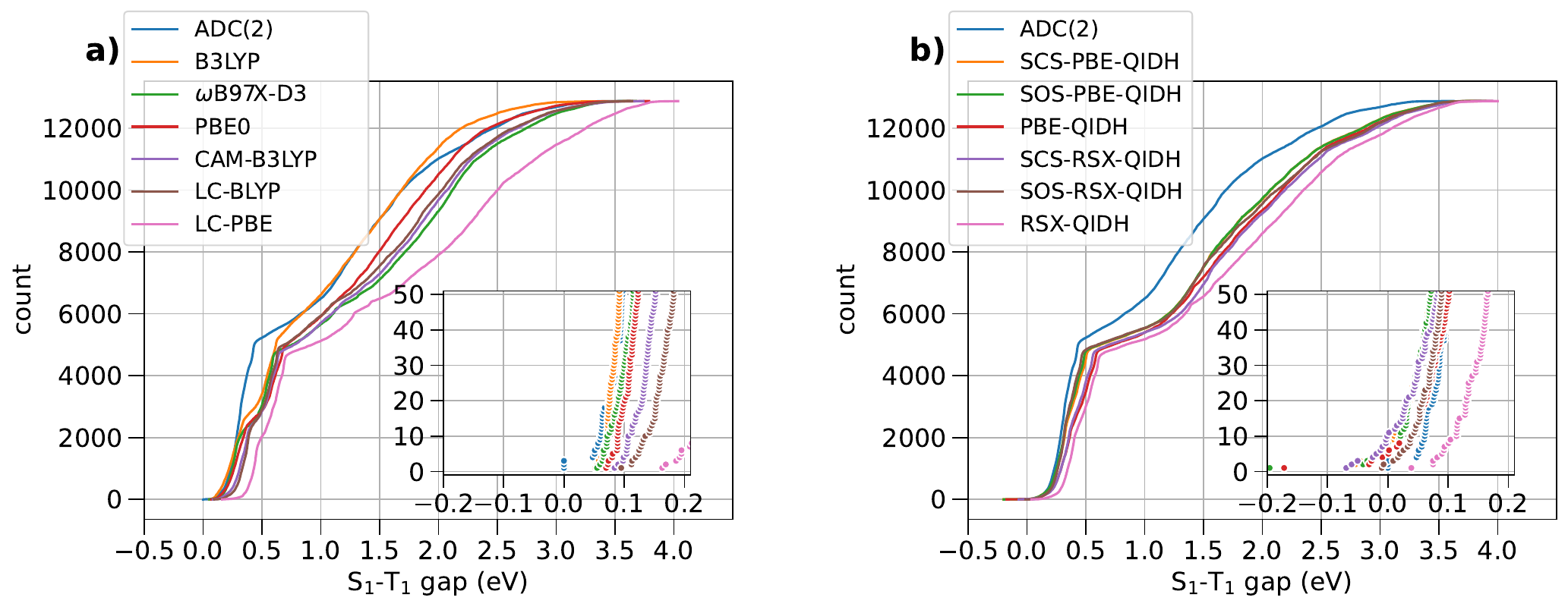}
\caption{Cumulative distribution of S$_1$-T$_1$ energy gaps of 12,880 bigQM7$\omega$ molecules
calculated with various {\it ab initio} methods. The inset shows the 
distribution in the range of $-$0.2 to 0.2 eV.}
\label{fig:cumulative}
\end{figure*}
In 2019, two independent studies confirmed STGs$<0$ in the prototypical cycl[3.3.3]azines---cyclazine\cite{de2019inverted} and heptazine\cite{ehrmaier2019singlet}---using time-dependent density functional theory (TDDFT) approximations, as well as a few correlated wavefunction methods. 
Since then, there has been a renewed interest in exploring the historically significant inverse-STG candidates: N-containing triangular molecules\cite{sanz2021negative, pollice2021organic, ricci2021singlet, sobolewski2021heptazine, aizawa2022delayed, tuvckova2022origin, bedogni2023shining, kim2024extension, blasco2024experimental} and 
non-alternant hydrocarbons\cite{terence2023symmetry, sandoval2023correlation,garner2024enhanced}. 
Besides these classes of molecules, Bedogni {\it et al.}\cite{bedogni2023shining} showed the possibility of STG$<0$ in C$_n$H$_n$N$_n$ aza-rings. 

Despite mounting computational evidence supporting the likelihood of Hund's rule violation, the credibility of negative STGs still attracts criticisms\cite{dreuw2023inverted}. This skepticism arises from the failure to account for the experimental conditions in computational modeling and the challenges posed by the large molecules for accurate {\it ab initio} calculations. More recently, Loos 
{\it et al.}\cite{loos2023heptazine} confirmed negative STGs in triangulene systems through composite excited states modeling and 
provided theoretical best estimates (TBEs).

The present study aims to report STGs calculated using a many-body method and double-hybrid density functional theory (dh-DFT) models for 12,880 small organic molecules with systematically varying compositions and structures.
Using this data, we verify the possibility of Hund's rule violation in the chemical space. 
Our workflow for data generation and the design of the module {\tt pymoldis} for querying the reported data 
is illustrated in FIG.~\ref{fig:workflow}. 
The rest of this article discusses the qualitative aspects of the data, their analysis, and the technical details of the calculations.

\section{Results and Discussions}
FIG.~\ref{fig:cumulative} shows the range spanned by the STGs of all the molecules in the bigQM7$\omega$ dataset determined using
various {\it ab initio} methods in the form of cumulative distributions. 
In both subplots (a and b) of this figure, ADC(2) is used as the reference theory for evaluating the accuracy of other methods. 
In the Supplementary Information (SI), we have discussed the accuracy of ADC(2) in combination with the def2-TZVP basis set for modeling STG of 10 triangle-shaped molecules using TBEs reported in a previous study\cite{loos2023heptazine}. 
Additionally, we have benchmarked the performance of the TDDFT version of SCS-PBE-QIDH within the Tamm--Dancoff approximation (TDA), see Tables~S1--S4. 

FIG.~\ref{fig:cumulative}a shows the cumulative count of STG of 12,880 molecules predicted with ADC(2) and selected hybrid-, long-range-corrected-DFT, and long-range-corrected-hybrid-DFT approximations. 
All the theoretical models featured in this plot predict positive STGs. 
Past studies\cite{ghosh2022origin,kondo2022singlet,tuvckova2022origin} have shown that explicit incorporation of electron correlation, for example, at the MP2-level  (second-order many-body perturbation theory) as in dh-DFT, is a requirement to predict STG$<0$. 
However, it has come as a surprise that at the ADC(2) level, which is the excited-state counterpart of MP2, none of the 12,880 molecules show a negative STG. 
Similar results from dh-DFT methods, along with their spin-component-scaled (SCS) and opposite-spin-component-scaled (SOS), are shown in FIG.~\ref{fig:cumulative}b. 
The zoomed-in inset shows the distribution of values from RSX-QIDH to shift towards the positive domain compared to ADC(2) values. 
Upon SCS/SOS corrections\cite{paez2021time}, the distribution is shifted slightly to the negative domain. 
At the dh-DFT level, SCS-PBE-QIDH, an accurate method for modeling STGs of triangulenes, a few molecules exhibit STG$<0$ eV; we give a detailed discussion of individual values later.

\begin{figure}[hb]
\centering
\includegraphics[width=\linewidth]{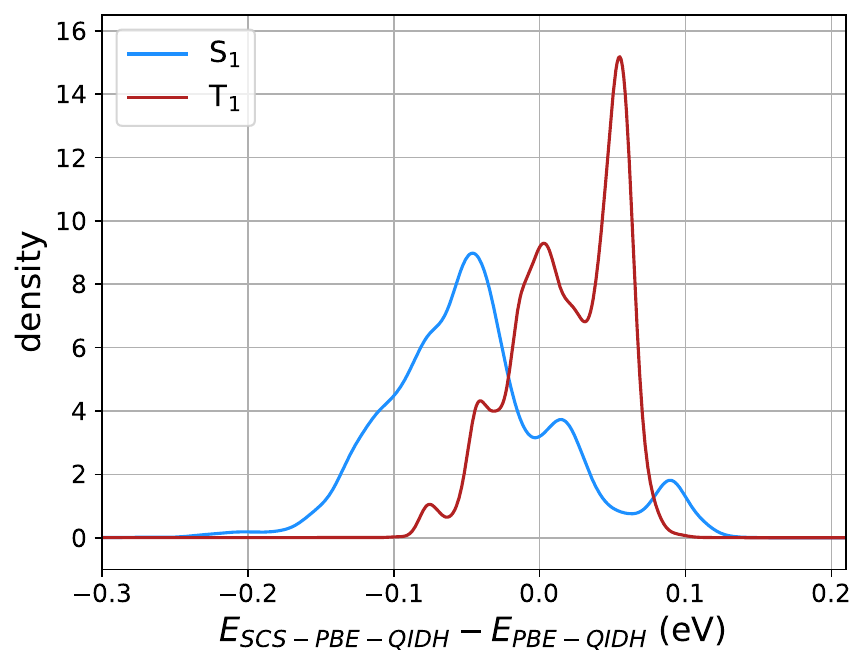}
\caption{
Probability density of the shift in S$_1$ and T$_1$ energies (in eV) for 12,880 molecules upon the inclusion of spin-component-scaling (SCS) in PBE-QIDH.}
\label{fig:scalingeffect}
\end{figure}

% % Fig. 1
% \begin{figure*}[ht]
%     \centering
%     \includegraphics[width=\linewidth]{SCS_SOS_checked.pdf}
%     \caption{
%     Probability density of the shift in S$_1$ and T$_1$ energies (in eV)
%     of 12,880 molecules
%     with the introduction of spin-component-scaling (SCS) and opposite-spin-component-scaling (SOS)
%     in PBE-QIDH and RSX-QIDH methods.
%     \INRED{Use this instead of Fig.3 and update the text accordingly}
%     }
%     \label{fig:SI_01}
% \end{figure*}

% \begin{figure*}[ht]
%     \centering
%     \includegraphics[width=\linewidth]{S1T1_gap_scatter_checked.png}
%     \caption{
%     Distribution of S$_1$ and T$_1$ energies of 12,880 molecules 
%     shown jointly with the S$_1$-T$_1$ gap from ADC(2) and SCS-PBE-QIDH methods. 
%     }
%     \label{fig:SI_02}
% \end{figure*}

FIG.~\ref{fig:scalingeffect} illustrates the effect of SCS corrections to PBE-QIDH predicted values of STG in the form of probability densities of the shift in S$_1$ and T$_1$ energies with the inclusion of SCS. 
Overall, the SCS corrections lower the S$_1$ energy while raising the T$_1$ values, illustrating why SCS-PBE-QIDH favors smaller STG values than the unscaled method, PBE-QIDH. 
Figure~S1 of the SI displays similar plots for SOS-PBE-QIDH, SCS-RSX-QIDH, and SOS-RSX-QIDH. 
While the SCS/SOS variants of RSX-QIDH shift the S$_1$ and T$_1$ energies further apart compared to the PBE-QIDH variants, previous benchmark studies\cite{loos2023heptazine,alipour2022any} have shown SCS-PBE-QIDH to be more accurate for modeling molecules with negative STGs. 
Hence, one can conclude that scaled-RSX-QIDH results in
more false-positive predictions ({\it i.e.} spurious predictions of STG$<0$ eV) than scaled-PBE-QIDH. 
Notably, the SCS/SOS corrections applied to the PBE-QIDH and RSX-QIDH dh-DFT methods are specifically tailored for the TD-DFT framework\cite{paez2021time}. 
Hence, STGs predicted using the $\Delta$SCF approach will be similar in unscaled and scaled DFT methods.

\begin{figure}[ht]
\centering
\includegraphics[width=\linewidth]{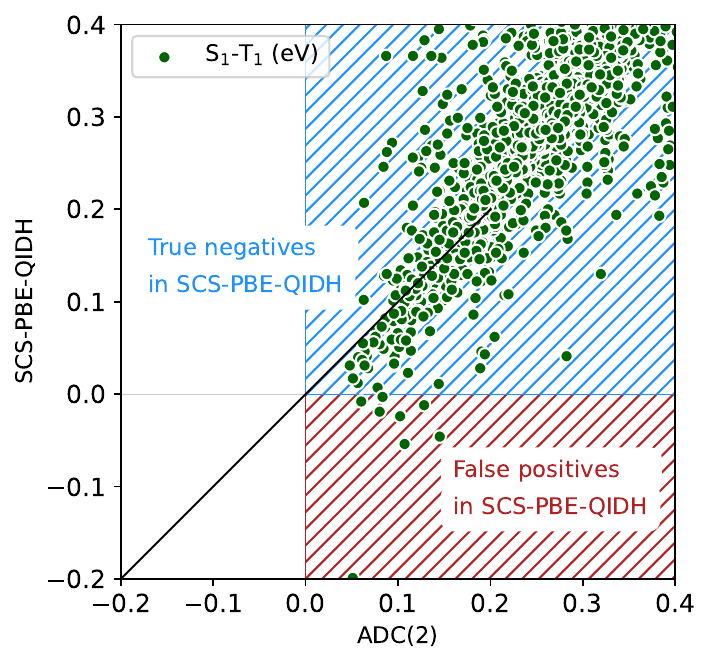}
\caption{
Scatterplot of S$_1$-T$_1$ energy gap $\le0.4$ eV from 
SCS-PBE-QIDH and ADC(2) for the bigQM7$\omega$ molecules.}
\label{fig:scatter}
\end{figure}

The scatterplot in FIG.~\ref{fig:scatter} offers a detailed view of the distribution of STG values around 0 eV. 
We find that the predominant entries shown in this plot are in the blue region denoted ``True negatives in SCS-PBE-QIDH", implying the TDDFT method agrees with ADC(2), both predicting these molecules as Hund's rule obeying systems with positive STGs. 
Molecules shown in the red region of FIG.~\ref{fig:scatter} are 
``False positives in SCS-PBE-QIDH" as these systems show STG$>0.04$ eV according to ADC(2), while in the TDDFT formalism they show STG$<0$ eV. 
As already highlighted in FIG.~\ref{fig:cumulative}, there are no ``True positives'' or ``False negatives'' as none of the 12,880 molecules in the bigQM7$\omega$ dataset exhibit a negative STG at the ADC(2) level. 

The boundary separating positives and negatives in FIG.~\ref{fig:scatter} is not sharp as both ADC(2) and SCS-PBE-QIDH have uncertainties $>0.1$ eV in their predictions (see benchmarks in the SI). 
Figure~S2 shows scatterplots of the joint distributions of STG with the S$_1$ and T$_1$ energies at both the SCS-PBE-QIDH and ADC(2) methods. 
At both levels of theories, one finds the small STG systems to have S$_1$ and T$_1$ energies in the range of 6--8 eV. 
Additionally, we find a molecule with S$_1$ and T$_1$ energy values in the 3-4 eV range to have a small STG at the SCS-PBE-QIDH level (see Figure~S2~c and Figure~S2~d). 
By querying the dataset using the {\tt pymoldis} module (see Figures~S3--S12), we found the corresponding molecule to be 2,6-dihydro-1H-pyridin-3-one (SMILES: {\tt O=C1CNCC=C1}), which is cyclohexenone with an N atom at the $\delta$-position. 

% \begin{figure*}[hbtp!]
% \centering
% \includegraphics[width=\linewidth]{SmallGapMols.pdf}
% \caption{
% Twelve molecules in the bigQM7$\omega$ dataset with the least S$_1$-T$_1$ gap at the SCS-PBE-QIDH level
% are shown per structure type. SMILES representation are given along with the
% S$_1$, T$_1$ and S$_1$-T$_1$ energies. 
% Predictions from ADC(2) are in parenthesis. All energies are in eV.
% }
% \label{fig:smallgaps}
% \end{figure*}

\begin{figure*}[hbtp!]
\centering
\includegraphics[width=\linewidth]{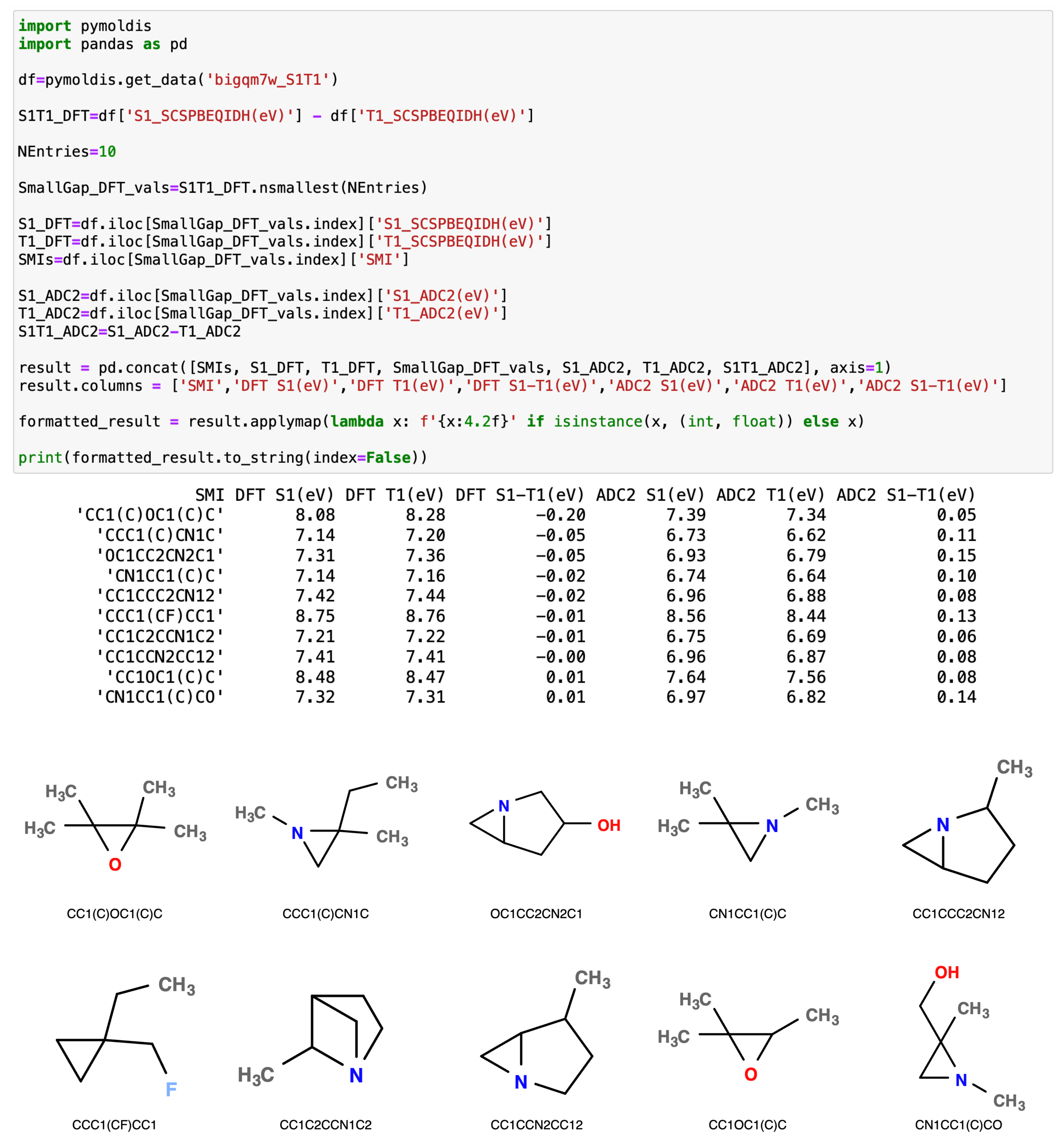}
\caption{
Data-mining the bigQM7$\omega$ dataset in {\tt pymoldis}\cite{pymoldis} 
to identify molecules with small S$_1$-T$_1$ gaps. For the ten molecules with the lowest gaps according to TDA-SCS-PBE-QIDH, SMILES strings and S$_1$/T$_1$ energies are displayed. The corresponding ADC(2) values are also given alongside. Molecular structures are displayed in cartoon format below.
See the SI for more examples of data-mining exercises.
}
\label{fig:smallgaps}
\end{figure*}

\begin{figure*}[hbtp!]
\centering
\includegraphics[width=\linewidth]{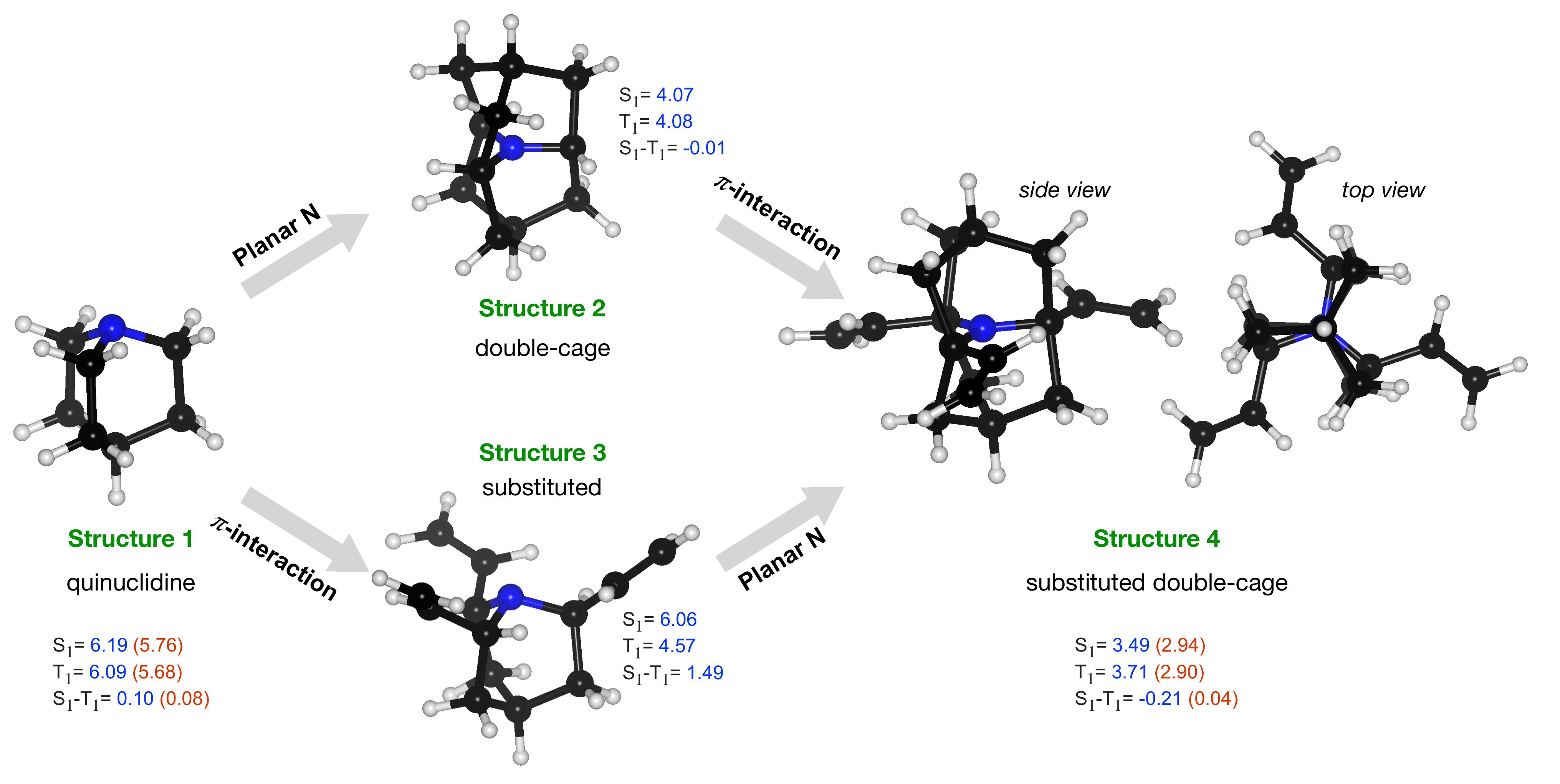}
\caption{
TDA/SCS-PBE-QIDH predictions of S$_1$, T$_1$ and S$_1$-T$_1$ energies (in eV) of 
1-azabicyclo[2.2.2]octane (aka quinuclidine) and its derivatives. For selected
structures, additionally,
ADC(2) values are in parentheses.
White, black, and blue atoms correspond to 
H, C, and N atoms, respectively.
}
\label{fig:cage}
\end{figure*}

To identify geometrical aspects common to the small STG molecules, we have queried the {\tt pymoldis} database for ten molecules
with the smallest STG according to TDA/SCS-PBE-QIDH. 
FIG.~\ref{fig:smallgaps} shows a screenshot of the query. 
A geometric moiety common to all N-containing systems is a substantial deviation of the bonding environment of the $sp^3$-N center from the ideal geometry. 
While for the ten benchmark triangular systems, the S$_1$ and T$_1$ excitation energies are $<3$ eV due to the possibility of a low-energy $n\rightarrow\pi^*$ transition, in the systems shown in FIG.~\ref{fig:smallgaps}, the S$_1$/T$_1$ excitations are of the $n\rightarrow\sigma^*$ type with excitation energies $>6$ eV. 
Nine of the ten molecules shown in FIG.~\ref{fig:smallgaps} contain a 3-membered heterocyclic ring, while one
is a fluorinated cyclopropane derivative; the S$_1$ and T$_1$ excitation energies of the latter are $>8$ eV, at both TDDFT and ADC(2) levels while the corresponding STG=$-$0.01 eV and 0.13 eV at TDDFT and ADC(2) levels, respectively.
The molecule corresponding to SMILES, {\tt CC1C2CCN1C2}, in FIG.~\ref{fig:smallgaps} contains a propellane-type cage with STG=$-$0.01 eV at the TDDFT level and an STG of 0.06 eV at the ADC(2) level. 
The bigQM7$\omega$ dataset comprises several such cage systems with small STGs. 
%Hence, in the following we explore the possibility to design a negative STG molecule starting with a cage system.

% \INBLUE{Even though there have not been any proven strategies to design systems with negative singlet-triplet gaps, few criteria could be concluded from previous studies. N containing systems,Conjugated pi systems, rigid planar geometry, poor overlap of hole and electron orbital. We had first attempted adding methyl groups to NH3. To make it more rigid we inserted isopropyl and tert-butyl groups.  }
Starting with a cage-type molecule with a strained N, we have 
explored the possibility of designing a molecule with the character 
of the lowest excitation as $n\rightarrow\pi^*$. 
We started with a symmetric cage system, quinuclidine, a [2.2.2]propellane with an axial CH group replaced by an N atom, Structure~1 in FIG.~\ref{fig:cage}. 
This molecule has an STG of 0.1 eV at the TDA-SCS-PBE-QIDH/def2-TZVP level. 
This small gap is because the excitation is primarily from non-bonding MO of N (highest occupied MO, HOMO) to C-H $\sigma^*$ of the cage (lowest unoccupied MO, LUMO); their corresponding densities show poor overlap leading to a small value of exchange integral, $K_{ar}$. 
Further, to arrive at a local-geometric environment of the N atom as in cycl[3.3.3]azines, we have introduced an additional cage to constrain the N-center to a plane (Structure~2 in FIG.~\ref{fig:cage}). 
This structure comprises perfectly co-planar C-N bonds resulting in degenerate S$_1$-T$_1$ levels. 
We have also modified quinuclidine by attaching three ethylene groups (Structure~3). In this structure, the S$_0\rightarrow$S$_1$ excitation has the character $n\rightarrow \pi^*$ (MO indices, $n$:52, which is the HOMO, and $\pi^*$:53--55) while the S$_0\rightarrow$T$_1$ excitation has the character $\pi\rightarrow\pi^*$ (MO indices, $\pi$:49--51)
with a large STG value of 1.5 eV. 
Finally, we combined structural modifications introduced in Structure~2 and Structure~3 to arrive at Structure~4 with a planar N interacting with $\pi$ moieties through space. 
Interestingly, this system resulted in an STG of $-$0.21 eV. 
To verify this prediction of an inverse-STG nature, we have also
performed ADC(2)/def2-TZVP calculations. 
For both Structure~1 and Structure~4, we find the magnitudes of the ADC(2) excitation energies to be lower than the DFT values. 
While both energies drop by a similar magnitude in the former, in Structure~4,  the energy of T$_1$ drops more than the S$_1$ energy, giving rise to a nearly zero STG at the ADC(2) level. 
This case study indicates that for molecules such as Structure~4, even some of the best double-hybrid DFT methods can spuriously predict a negative STG; hence, one must consider many-body methods such as ADC(2) as a baseline theory.

\begin{figure}[ht]
\centering
\includegraphics[width=\linewidth]{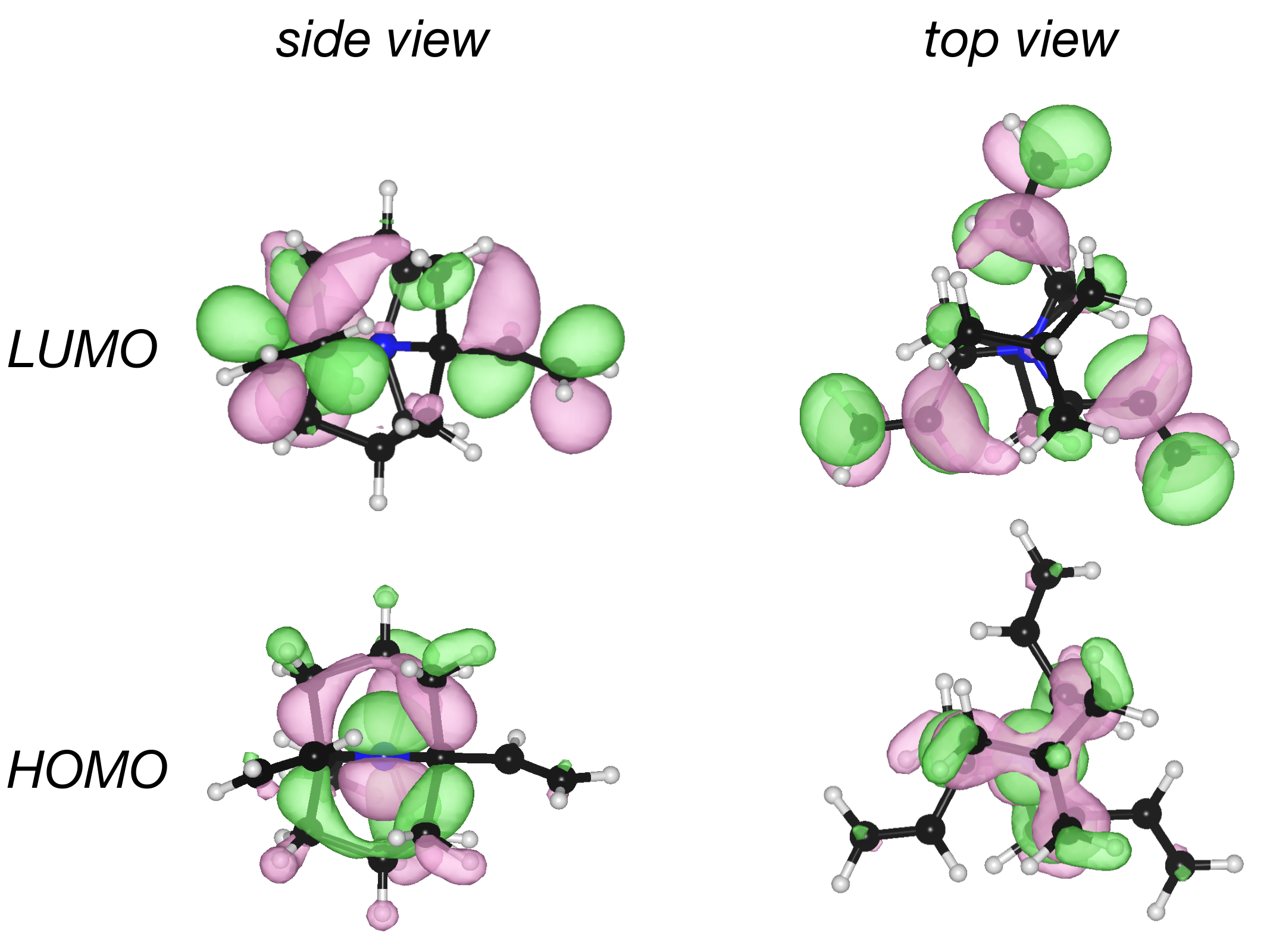}
\caption{
Plots of HOMO and LUMO of the ethylene substituted double cage derivative
of quinuclidine, Structure~4, shown in FIG.~\ref{fig:cage}.
}
\label{fig:MOs}
\end{figure}

While Structure~4 is a minimum on the potential energy surface as verified through vibrational frequency analysis, we do not expect the system to be relevant to the thermally activated delayed fluorescence (TADF) applications\cite{de2019inverted,li2022organic,pollice2021organic,wang2023inverted,won2023inverted}. 
On the other hand, it is a compelling computational chemistry 
exercise to modulate a molecule's STG by chemical modifications. 
Hence, even though Structure~4 seems to be yet another false positive
in the search for a
Hund's rule-violating molecule, we have examined it further. 
We inspected the shape of the MOs involved in the S$_1$ and T$_1$ excitations and found the excitations to be primarily HOMO$\rightarrow$LUMO type. 
These MOs are on display in FIG.~\ref{fig:MOs}, from which we visually conclude that the
densities of HOMOs and LUMOs do not overlap. 
For a more quantitative analysis, we calculated the $\Lambda$-index\cite{peach2008excitation} defined as $\int d{\bf r}\, |\phi_a({\bf r})||\phi_r({\bf r})|$ using Multiwfn\cite{lu2012multiwfn}, and obtained the values: 0.37 and 0.40 for the S$_1$ and the T$_1$ states, respectively. The $\Lambda$-index quantifies the degree of overlap between hole and electron in S$_0\rightarrow$S$_1$ and S$_0\rightarrow$T$_1$ excitations. In comparison, for cyclazine and heptazine, the values of $\Lambda$ for the S$_1$/T$_1$ states are 0.49/0.49 and 0.50/0.51, respectively.

\section{Computational details}
The bigQM7$\omega$ dataset \cite{kayastha2022resolution,kayastha2021Supplementary} features 
12,880 molecules with up to 7 CONF atoms with equilibrium geometries determined with the $\omega$B97X-D DFT method and the def2-TZVP basis set. Molecular graphs of bigQM7$\omega$ molecules, encoded as SMILES, were sourced from the GDB11\cite{fink2005virtual} chemical space, which contains several common molecules, such as acetic acid and benzoic acid, which the subsequent databases GDB13\cite{blum2009970} and GDB17\cite{ruddigkeit2012enumeration} filtered out.
Consequently, bigQM7$\omega$ offers over thrice the number of molecules with up to 7 CONF atoms than the QM9 dataset\cite{ramakrishnan2014quantum} derived from GDB17. 
The geometries of the bigQM7$\omega$ molecules were 
optimized\cite{kayastha2022resolution} using the connectivity 
preserving geometry optimization (ConnGO) workflow to prevent 
covalent bond rearrangements during 
geometry optimization\cite{senthil2021troubleshooting}. 
Further, vibrational 
frequency analysis at the $\omega$B97X-D/def2-TZVP level
verified each structure in bigQM7$\omega$ to be an energy minimum. 
Along with the minimum energy geometries, bigQM7$\omega$
offers several ground-state properties ranging from partial charges to 
thermochemistry energies along with excited state properties\cite{kayastha2021Supplementary}.

For all 12,880 molecules in bigQM7$\omega$, we performed single-point vertical excited state 
calculations of the S$_1$ and T$_1$ energies using 12 DFT methods:
PBE0 \cite{adamo1999toward}, B3LYP \cite{stephens1994ab}, CAM-B3LYP \cite{yanai2004new}, 
$\omega$B97X-D3\cite{chai2008long}, LC-BLYP \cite{tawada2004long}, LC-PBE \cite{iikura2001long},
PBE-QIDH \cite{bremond2014communication}, SCS-PBE-QIDH, SOS-PBE-QIDH\cite{paez2021time}, 
RSX-QIDH\cite{bremond2018range}, SCS-RSX-QIDH \cite{paez2021time} and SOS-RSX-QIDH \cite{paez2021time}. 
We also calculated S$_1$ and T$_1$ energies 
using the correlated excited state method: second-order
algebraic diagrammatic construction, ADC(2). 
The accuracy of ADC(2) and SCS-PBE-QIDH in combination with other settings is evaluated in the SI using previously reported\cite{loos2023heptazine} TBEs of 10 triangular systems as references. For this purpose, we 
performed geometry optimization of the triangular molecules 
using the $\omega$B97X-D3 DFT method with {\tt tightscf} and {\tt tightopt} keywords
in combination with the def2-TZVP basis set. 
Minimum energy structures and S$_1$/T$_1$ energies of the bigQM7$\omega$ can be queried using the {\tt pymoldis} module presented in this study (See Figure~S08 and Figure~S12 in the SI).

We performed ADC(2) calculations using QChem~6.0.2 and DFT calculations using ORCA~5.0.4\cite{neese2012orca,neese2018software}. 
With in TDDFT, we calculated twelve energy eigenvalues---six singlets and six triplets---which we sorted separately to extract S$_1$ and T$_1$ (lowest excited triplet) TDDFT excitation energies. 
In all calculations, we used the resolution-of-identity (RI) approximation\cite{vahtras1993integral,kendall1997impact}. 
In DFT calculations, we used the `chain-of-spheres' (COS) algorithm for exchange integrals (RIJCOSX). In dh-DFT calculations, 
we used the universal fitting auxiliary basis sets by Weigend \cite{weigend2006accurate} (denoted {\tt def2/J}) along with the def2-TZVP/C and aug-cc-pVTZ/C basis sets for the orbital basis sets def2-TZVP, and aug-cc-pVTZ, respectively.

\section{Conclusions}
We have probed the violation of Hund's rule in the chemical space of about 13,000 small organic molecules
with up to 7 atoms of C/O/N/F. 
We performed high-throughput calculations of excited states with various DFT methods and the more accurate theory ADC(2). 
We selected these methods based on their accuracy in benchmark calculations and compared them with previously reported theoretical best estimates for the STGs of ten triangular molecules. 
ADC(2) with a triple-zeta basis set provides an effective cost-accuracy trade-off for generating large-scale data. 
Further, this method has been shown\cite{loos2023heptazine} to agree with composite excited state methods for predicting ${\rm STG}<0$ with an average error of $<0.05$ eV. 
The critical result of the present study is that Hund's rule\cite{kutzelnigg1996hund} prevails across thousands of organic molecules with systematically varying structures covering almost all prototypical small organic molecules. 

The data presented in this study is importable in Python code for data mining endeavors. Using this infrastructure, we identified molecules with vanishing STG, some of which have negative values at the SCS-PBE-QIDH level while not violating Hund's rule as per ADC(2) predictions. A common geometric feature of these molecules was a substantial deviation of an N-atom from the typical $sp^3$ environment with both singlet and triplet excitations showing the $n\rightarrow \sigma^*$ character and potentially vanishing exchange interaction integral between the MOs involved in excitation. 

We have selected a cage structure and attached ethylene groups to mimic the environment of the N atom as in the well-known cases of cycl[3.3.3]azines. The corresponding MOs involved in S$_1$/T$_1$ excitations exhibit characteristics seen in previously studied triangular negative-STG systems. Upon further scrutiny, we showed this molecule as a conventional molecule obeying Hund's rule. Yet, introducing a polarizable environment in this system through donor-acceptor groups may selectively stabilize the S$_1$ over T$_1$\cite{de2019inverted,garner2024enhanced}. 
In this study, we did not investigate the practical utility of the small molecules studied in the context of TADF. 
Such exploration necessitates meticulous consideration of adiabatic effects on a case-by-case basis, a task that exceeds the scope of our present investigation. 
The present study demonstrates that a data-driven approach allows for gaining insight into the molecular structural factors that can quench the singlet-triplet energy gap. 
We offered evidence that the chemical space of small closed-shell 
organic molecules lack geometric and electronic structural 
necessary for a negative S$_1$-T$_1$ energy gap.

Theoretical studies have identified only a few molecular fingerprints to favor negative STGs. Dynamic spin polarization, attributed to double excitation effects involving frontier orbitals, has emerged as a potential mechanism to induce a negative STG \cite{kollmar1978violation,pollice2024rational,drwal2023role}. 
While a quantitative relationship exists between molecular structural features and zero STG, a corresponding structure-property relation for negative STG remains elusive. 
Introducing functional groups is one promising avenue for designing large synthetically tractable molecules with negative STG. 
For theoretical explorations, our research highlights the limitations of using DFT methods, which can result in false positives. 
Consequently, there is a pressing need for efficient strategies to accelerate predictions using correlated wavefunction methods. 
For large-scale investigations, data-driven modeling can complement first-principles modeling combined with inverse-design strategies, such as those based on genetic algorithms\cite{nigam2024artificial, gupta2021data}.

\section{Supplementary Information}
Contains the following: 
Assessment of S$_1$-T$_1$ gaps from ADC(2) and SCS-PBE-QIDH for 10 triangular molecules.
Table~S1 compares S$_1$/T$_1$ energetics calculated using the ADC(2) method 
with theoretical best estimates.
Table~S2 compares the S$_1$/T$_1$ energetics predicted by TDDFT and TDA.
Table~S3 contains ADC(2) and TDA energies calculated using
DFT-level geometries. Table~S4 provides various error 
metrics for ADC(2) and SCS-PBE-QIDH predicted S$_1$/T$_1$ energetics.
Figure~S1 illustrates the shifts in S$_1$ and T$_1$ due to SCS/SOS 
corrections to PBE-QIDH and RSX-QIDH methods. 
Figure~S2 shows a scatterplot of S$_1$ and T$_1$ energies with S$_1$-T$_1$ gaps.
Figures~S3–S12 offer screenshots of data mining exercises.
Minimum energy geometries of Structures 1--4 in FIG.~\ref{fig:cage} are also listed.
Sample Python notebooks and further details are available at {\tt https://github.com/moldis-group/pymoldis}\cite{pymoldis}.

\section{Data Availability}
The data that support the findings of this study are
within the article and its supplementary material.

\section{Acknowledgments}
We acknowledge the support of the Department of Atomic Energy, Government
of India, under Project Identification No.~RTI~4007. 
All calculations have been performed using the Helios computer cluster, 
which is an integral part of the MolDis Big Data facility, TIFR Hyderabad \href{http://moldis.tifrh.res.in}{(http://moldis.tifrh.res.in)}.

\section{Author Declarations}

\subsection{Author contributions}
\noindent 
{\bf AM}: 
Conceptualization (equal); 
Analysis (equal); 
Data collection (equal); 
Writing (equal);
Revision (equal).
{\bf RR}: Conceptualization (equal); 
Analysis (equal); 
Data collection (equal); 
Funding acquisition; 
Project administration and supervision; 
Resources; 
Writing (equal);
Revision (equal).

\subsection{Conflicts of Interest}
The authors have no conflicts of interest to disclose.
%\singlespacing

\section*{References}
\bibliography{ref} 
\end{document}

\subsection{Assessment of methods for data-generation}
For calculating the energies of the S$_1$ and T$_1$ excited states 
of all molecules in the bigQM7$\omega$ dataset,  we have evaluated the 
cost-accuracy trade-off in various approximations stemming from the choice of the basis set, accuracy of geometries, approximation of 
molecular integrals with resolution-of-identity, and the Tamm-Dancoff approximation (TDA) to time-dependent density functional theory (TDDFT). To this end, we have selected ten triangular molecules from a previous study\cite{loos2023heptazine} with the theoretical best estimates (TBEs) for the energies of S$_1$ and T$_1$ determined using frozen-core CCSD(T)/cc-pVTZ geometries. From the same study, we collected energies from ADC(2) and the TDDFT-SCS-PBE-QIDH methods
using the aug-cc-pVTZ basis set. With these values as references, we 
evaluated the accuracy of ADC(2) and TDA using the smaller basis set def2-TZVP.

For all benchmark systems, S$_1$ and T$_1$ energies, along with the S$_1$-T$_1$ gaps determined with various theoretical methods, are collected in Table~S1, Table~S2, and Table~S3. The accuracies of different methods are quantified via the error metrics, mean signed error (MSE), mean absolute error (MAE), standard deviation of the error (SDE),  minimum error (minE), and maximum error (maxE) in Table~S4. We evaluate these metrics using the aforestated TBEs\cite{loos2023heptazine} as the reference. To understand the impact of the choice of geometries, we have performed geometry optimization of these ten molecules at the $\omega$B97X-D3/def2-TZVP level. 

From Table~S1, we find that when using frozen-core CCSD(T)/cc-pVTZ geometries, ADC(2)/def2-TZVP predicts similar energies as ADC(2)/aug-cc-pVTZ. However, the error metrics in Table~S4 suggest that for all S$_1$-T$_1$ gaps, the SDE drops from 0.034 eV to 0.011 eV, indicating the smaller basis set def2-TZVP to perform better in the ADC(2) calculations. We also note similar improvement for S$_1$ and T$_1$ energies. This trend is in line with the observation made in an earlier study\cite{curtis2023novel} that ADC2/aug-cc-pVDZ yields a more accurate S$_1$-T$_1$ gap for triangular molecules than ADC(2)/aug-cc-pVTZ due to a favorable error cancellation. ADC(2)/def2-TZVP values determined using $\omega$B97X-D3/def2-TZVP geometries (see Table~S3) very closely resemble ADC(2)/def2-TZVP values determined using CCSD(T)/cc-pVTZ geometries (see Table~S1). Further, while the change in geometry has a small influence on the error metrics for S$_1$ and T$_1$ energies, due to the cancellation of effects, the metrics for the S$_1$-T$_1$ gap are less influenced. 

We also compare the TDDFT/SCS-PBE-QIDH/aug-cc-pVTZ (determined using CCSD(T)/cc-pVTZ geometries) results from Ref.~1,
to TDA analog in Table~S2. In this Table, we have also reported TDDFT values calculated by us using geometries from Ref.~1 and find our results to agree with previously reported values precisely, assuring that our computational settings are consistent with the previous study. Moving from TDDFT to TDA, the largest effect is seen for molecule 6. At the TDDFT level, this system's S$_1$-T$_1$ gap is $-$0.32 eV, which increases to $-$0.03 eV at the TDA level. However, we find the TDA value better to approximate the TBE value of $-$0.07 eV from Table~S1. For the S$_1$-T$_1$ gap, TDDFT values of MSE, MAE, and SDE are $-$0.033, 0.055, and 0.081 eV, respectively, which drop to 0.034, 0.034, and 0.012 eV when using TDA. The SDE of  TDA is 8-fold smaller than that of TDDFT, indicating smaller non-systematic errors in the predictions.

The TDA calculations when performed using def2-TZVP basis set, in combinaton with $\omega$B97X-D3/def2-TZVP geometries (Table~S3),
result in increasing MSE and MAE for S$_1$ and T$_1$ energies.   
However, due to the cancellation of effects, the MAE and SDE for the S$_1$-T$_1$ gap are smaller than TDDFT/aug-cc-pVTZ and TDA/aug-cc-pVTZ (see Table~S4); the latter two using CCSD(T)/cc-pVTZ geometries from Ref.~1. 

Hence, combining the favorable effects of TDA and a smaller basis set, we have used TDA-TDDFT/def2-TZVP in combination with various DFT approximations to generate the excited state energetics of the bigQM7$\omega$ dataset.

%TABLE 1
%\clearpage

\begin{table*}[ht]
\begin{threeparttable}
\centering
\caption{Energies of the
S$_1$ and T$_1$ states with respect to the S$_0$
ground state
along with the singlet-triplet gap, S$_1$-T$_1$,
of 10 triangular benchmark systems reported in Ref.~1.  
ADC(2)/def2-TZVP results are compared with 
theoretical best estimate (TBE) and ADC(2)/aug-cc-pVTZ values from Ref.~1.
%\RRef{loos2023heptazine}.
All values are in eV and
\# indicates
compound number. In all calculations, we used 
geometries determined with the CCSD(T)/cc-pVTZ method reported in Ref.~1.
 }
\small\addtolength{\tabcolsep}{1.2pt}
\begin{tabular}[t]{l llll llll lll}
\hline
\multicolumn{1}{l}{\#}   &
\multicolumn{4}{l}{TBE$^a$} & 
\multicolumn{4}{l}{ADC(2)/aug-cc-pVTZ$^a$} & 
\multicolumn{3}{l}{ADC(2)/def2-TZVP$^b$} \\
\cline{2-4} \cline{6-8} \cline{10-12} 
\multicolumn{1}{l}{} &
\multicolumn{1}{l}{S$_1$} &
\multicolumn{1}{l}{T$_1$} &
\multicolumn{1}{l}{S$_1$-T$_1$} &
\multicolumn{1}{l}{} &
\multicolumn{1}{l}{S$_1$} &
\multicolumn{1}{l}{T$_1$} &
\multicolumn{1}{l}{S$_1$-T$_1$} &
\multicolumn{1}{l}{} &
\multicolumn{1}{l}{S$_1$} &
\multicolumn{1}{l}{T$_1$} &
\multicolumn{1}{l}{S$_1$-T$_1$} \\
\hline 
1 & 2.717 & 2.936 & -0.219 & & 2.675 & 2.921 & -0.246 & & 2.665 & 2.915 & -0.250 \\ 
2 & 0.979 & 1.111 & -0.131 & & 1.001 & 1.138 & -0.137 & & 1.001 & 1.139 & -0.138 \\ 
3 & 1.562 & 1.663 & -0.101 & & 1.551 & 1.664 & -0.113 & & 1.548 & 1.665 & -0.117 \\ 
4 & 2.177 & 2.296 & -0.119 & & 2.159 & 2.298 & -0.139 & & 2.153 & 2.295 & -0.142 \\ 
5 & 2.127 & 2.230 & -0.103 & & 2.098 & 2.225 & -0.127 & & 2.093 & 2.225 & -0.132 \\ 
6 & 0.833 & 0.904 & -0.071 & & 0.851 & 0.945 & -0.094 & & 0.854 & 0.950 & -0.096 \\ 
7 & 0.693 & 0.735 & -0.042 & & 0.708 & 0.782 & -0.074 & & 0.714 & 0.791 & -0.078 \\ 
8 & 0.554 & 0.583 & -0.029 & & 0.565 & 0.635 & -0.070 & & 0.575 & 0.646 & -0.071 \\ 
9 & 1.264 & 1.463 & -0.199 & & 1.274 & 1.488 & -0.214 & & 1.271 & 1.487 & -0.216 \\ 
10 & 1.522 & 1.827 & -0.305 & & 1.639 & 2.074 & -0.435 & & 1.526 & 1.840 & -0.314 \\
\hline
\end{tabular}
\begin{tablenotes}
 \item \footnotesize{$^a$ From Ref.~1.}
 \item \footnotesize{$^b$ This work.}
 \end{tablenotes}
\end{threeparttable}
\label{tab:W1}
\end{table*}%

%TABLE 2

% \clearpage
\begin{table*}[ht]
\begin{threeparttable}
\centering
\caption{Energies of the
S$_1$ and T$_1$ states with respect to the S$_0$
ground state along with the singlet-triplet
gap, S$_1$-T$_1$,
of 10 triangular benchmark systems
reported in Ref.~1.  
All energies were determined using
TDDFT or its Tamm–Dancoff approximation (TDA) employing
the SCS-PBE-QIDH double-hybrid DFT method
and the aug-cc-pVTZ basis set.
All values are in eV and
\# indicates
compound number.
In all calculations, we used 
geometries determined with the CCSD(T)/cc-pVTZ method reported in Ref.~1.
 }
\small\addtolength{\tabcolsep}{1.2pt}
\begin{tabular}[t]{l llll llll lll}
\hline
\multicolumn{1}{l}{\#}   &
\multicolumn{4}{l}{TDDFT/aug-cc-pVTZ$^a$} & 
\multicolumn{4}{l}{TDDFT/aug-cc-pVTZ$^b$} & 
\multicolumn{3}{l}{TDA/aug-cc-pVTZ$^b$} \\
\cline{2-4} \cline{6-8} \cline{10-12} 
\multicolumn{1}{l}{} &
\multicolumn{1}{l}{S$_1$} &
\multicolumn{1}{l}{T$_1$} &
\multicolumn{1}{l}{S$_1$-T$_1$} &
\multicolumn{1}{l}{} &
\multicolumn{1}{l}{S$_1$} &
\multicolumn{1}{l}{T$_1$} &
\multicolumn{1}{l}{S$_1$-T$_1$} &
\multicolumn{1}{l}{} &
\multicolumn{1}{l}{S$_1$} &
\multicolumn{1}{l}{T$_1$} &
\multicolumn{1}{l}{S$_1$-T$_1$} \\
\hline 
1 & 2.770 & 2.987 & -0.217 & & 2.770 & 2.986 & -0.216 & &    2.845 &  3.055&  -0.210\\
2 & 1.039 & 1.163 & -0.124 & & 1.037 & 1.161 & -0.124 & &    1.112 &  1.196&  -0.084\\
3 & 1.621 & 1.685 & -0.064 & & 1.620 & 1.685 & -0.065 & &    1.696 &  1.749&  -0.053\\
4 & 2.239 & 2.340 & -0.101 & & 2.238 & 2.339 & -0.101 & &    2.317 &  2.405&  -0.088\\
5 & 2.188 & 2.245 & -0.057 & & 2.186 & 2.243 & -0.057 & &    2.264 &  2.331&  -0.067\\
6 & 0.881 & 1.201 & -0.320 & & 0.879 & 1.202 & -0.323 & &    0.959 &  0.986&  -0.027\\
7 & 0.728 & 0.825 & -0.097 & & 0.726 & 0.824 & -0.098 & &    0.809 &  0.808&   0.001\\
8 & 0.574 & 0.673 & -0.099 & & 0.573 & 0.672 & -0.099 & &    0.658 &  0.647&   0.011\\
9 & 1.305 & 1.538 & -0.233 & & 1.305 & 1.538 & -0.233 & &    1.398 &  1.567&  -0.169\\
10 & 1.566 & 1.906 & -0.340 & & 1.566 & 1.906 & -0.340& &    1.659 &  1.947&  -0.288\\

\hline
\end{tabular}
\begin{tablenotes}
\item \footnotesize{$^a$ From Ref.~1.}
\item \footnotesize{$^b$ This work.}
\end{tablenotes}
\end{threeparttable}
\label{tab:W1}
\end{table*}%

%TABLE 3

\begin{table*}[ht]
%\begin{threeparttable}
\centering
\caption{ADC(2) and TDA-SCS-PBE-QIDH 
energies of the
S$_1$ and T$_1$ states with respect to the S$_0$
ground state along with the singlet-triplet
gap, S$_1$-T$_1$,
of 10 triangular benchmark systems
reported in Ref.~1.  
All values are in eV and
\# indicates
compound number.
All calculations were performed using the def2-TZVP basis set, and geometries 
calculated with the $\omega$B97X-D3/def2-TZVP method.
 }
\small\addtolength{\tabcolsep}{1.2pt}
\begin{tabular}[t]{l llll lll}
\hline
\multicolumn{1}{l}{\#}   &
\multicolumn{4}{l}{ADC(2)/def2-TZVP} & 
\multicolumn{3}{l}{TDA/def2-TZVP} \\
\cline{2-4} \cline{6-8} 
\multicolumn{1}{l}{} &
\multicolumn{1}{l}{S$_1$} &
\multicolumn{1}{l}{T$_1$} &
\multicolumn{1}{l}{S$_1$-T$_1$} &
\multicolumn{1}{l}{} &
\multicolumn{1}{l}{S$_1$} &
\multicolumn{1}{l}{T$_1$} &
\multicolumn{1}{l}{S$_1$-T$_1$} \\
\hline 
1    & 2.691  &  2.935  & -0.244  & & 2.865  &  3.070  & -0.205 \\
2    & 1.021  &  1.158  & -0.137  & & 1.135  &  1.219  & -0.084 \\
3    & 1.569  &  1.684  & -0.115  & & 1.716  &  1.770  & -0.054 \\
4    & 2.175  &  2.313  & -0.139  & & 2.336  &  2.420  & -0.084 \\
5    & 2.115  &  2.243  & -0.128  & & 2.283  &  2.350  & -0.067 \\
6    & 0.884  &  0.982  & -0.098  & & 0.995  &  1.027  & -0.032 \\
7    & 0.752  &  0.830  & -0.078  & & 0.857  &  0.860  & -0.003 \\
8    & 0.620  &  0.689  & -0.069  & & 0.715  &  0.706  &  0.009 \\
9    & 1.265  &  1.480  & -0.215  & & 1.390  &  1.558  & -0.168 \\
10   & 1.518  &  1.828  & -0.310  & & 1.648  &  1.931  & -0.283 \\
\hline
\end{tabular}
% \begin{tablenotes}
% \item \footnotesize{$^a$ Using  geometries determined at the 
% CCSD(T)/cc-pVTZ, from Ref.~1.}
% \item \footnotesize{$^b$ Using $\omega$B97X-D3/def2-TZVP geometry, this work.}
% \end{tablenotes}
%\end{threeparttable}
\label{tab:W1}
\end{table*}%

%TABLE 4

\begin{table*}[ht]
\begin{threeparttable}
\centering
\caption{Error metrics for predicting the
S$_1$ \& T$_1$ energetics of 10 triangular benchmark systems
reported in Ref.~1. Values are reported for 
various theoretical 
methods in comparison with the theoretical best estimates from Ref.~1.
MSE: mean signed error,
MAE: mean absolute error,
SDE: standard deviation of the error,
minE: minimal error,
maxE: maximal error are in eV.
 }
\small\addtolength{\tabcolsep}{1.2pt}
\begin{tabular}[t]{l l rrrrr}
\hline
\multicolumn{1}{l}{Method}   &
\multicolumn{1}{l}{Energy} & 
\multicolumn{1}{l}{MSE} & 
\multicolumn{1}{l}{MAE} &
\multicolumn{1}{l}{SDE} &
\multicolumn{1}{l}{minE} &
\multicolumn{1}{l}{maxE} \\
\hline 
ADC(2)/aug-cc-pVTZ$^a$  &  S$_1$      &    0.009 &    0.029 &    0.041 &   -0.042 &    0.117 \\
                   &  T$_1$       &    0.042 &    0.046 &    0.072 &   -0.015 &    0.247\\
                   &  S$_1$-T$_1$ &   -0.033 &    0.033 &    0.034 &   -0.130 &   -0.006\\
                   &  mean        &    0.006 &    0.036 &    0.049 &   -0.062 &    0.119\\
ADC(2)/def2-TZVP$^b$  &  S$_1$      &   -0.003 &    0.022 &    0.025 &   -0.052 &    0.022\\
                   &  T$_1$       &    0.021 &    0.026 &    0.027 &   -0.021 &    0.063\\
                   &  S$_1$-T$_1$ &   -0.024 &    0.024 &    0.011 &   -0.042 &   -0.007\\
                   &  mean        &   -0.002 &    0.024 &    0.021 &   -0.038 &    0.026\\
ADC(2)/def2-TZVP$^c$  &  S$_1$      &    0.018 &    0.027 &    0.031 &   -0.026 &    0.066\\
                   &  T$_1$       &    0.040 &    0.040 &    0.038 &   -0.001 &    0.106\\
                   &  S$_1$-T$_1$ &   -0.021 &    0.021 &    0.011 &   -0.040 &   -0.005\\
                   &  mean        &    0.012 &    0.029 &    0.027 &   -0.022 &    0.056\\
TDDFT-SCS-PBE-QIDH/aug-cc-pVTZ$^a$  &  S$_1$      &    0.048 &    0.048 &    0.013 &    0.020 &    0.062\\
                   &  T$_1$       &    0.082 &    0.082 &    0.076 &    0.015 &    0.297\\
                   &  S$_1$-T$_1$ &   -0.033 &    0.055 &    0.081 &   -0.249 &    0.046\\
                   &  mean        &    0.032 &    0.062 &    0.057 &   -0.071 &    0.135\\
TDA-SCS-PBE-QIDH/aug-cc-pVTZ$^b$  &  S$_1$      &    0.129 &    0.129 &    0.011 &    0.104 &    0.140\\
                   &  T$_1$       &    0.094 &    0.094 &    0.018 &    0.064 &    0.120\\
                   &  S$_1$-T$_1$ &    0.034 &    0.034 &    0.012 &    0.009 &    0.048\\
                   &  mean        &    0.086 &    0.086 &    0.014 &    0.059 &    0.103\\
TDA-SCS-PBE-QIDH/def2-TZVP$^c$  &  S$_1$      &    0.151 &    0.151 &    0.013 &    0.126 &    0.164\\
                   &  T$_1$       &    0.116 &    0.116 &    0.011 &    0.095 &    0.134\\
                   &  S$_1$-T$_1$ &    0.035 &    0.035 &    0.010 &    0.014 &    0.047\\
                   &  mean        &    0.101 &    0.101 &    0.011 &    0.078 &    0.115\\
\hline
\end{tabular}
\begin{tablenotes}
\item \footnotesize{$^a$ Using CCSD(T)/cc-pVTZ geometries from Ref.~1.}
\item \footnotesize{$^a$ This work, using CCSD(T)/cc-pVTZ geometries from Ref.~1.}
\item \footnotesize{$^c$ Using $\omega$B97X-D3/def2-TZVP geometries, this work.}
\end{tablenotes}
\end{threeparttable}
\label{tab:W1}
\end{table*}%

%\INBLUE{Add back the figures of scatter plot and scaling plot}
% Fig. 1
\begin{figure*}[ht]
    \centering
    \includegraphics[width=\linewidth]{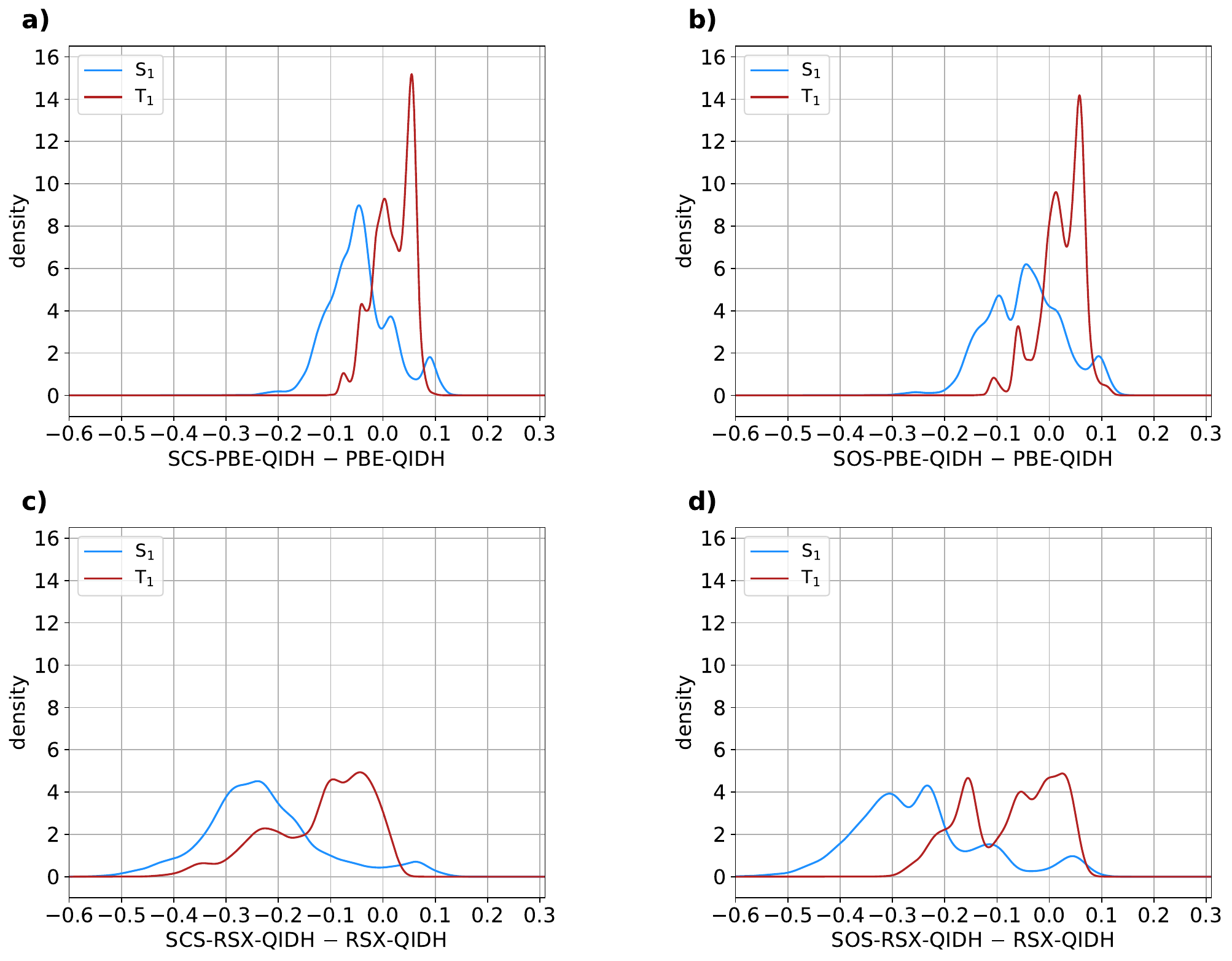}
    \caption{
    Probability density of the shift in S$_1$ and T$_1$ energies (in eV)
    of 12,880 molecules
    with the introduction of spin-component-scaling (SCS) and opposite-spin-component-scaling (SOS)
    in PBE-QIDH and RSX-QIDH methods.}
    \label{fig:SI_01}
\end{figure*}

% Fig. 2
\begin{figure*}[ht]
    \centering
    \includegraphics[width=\linewidth]{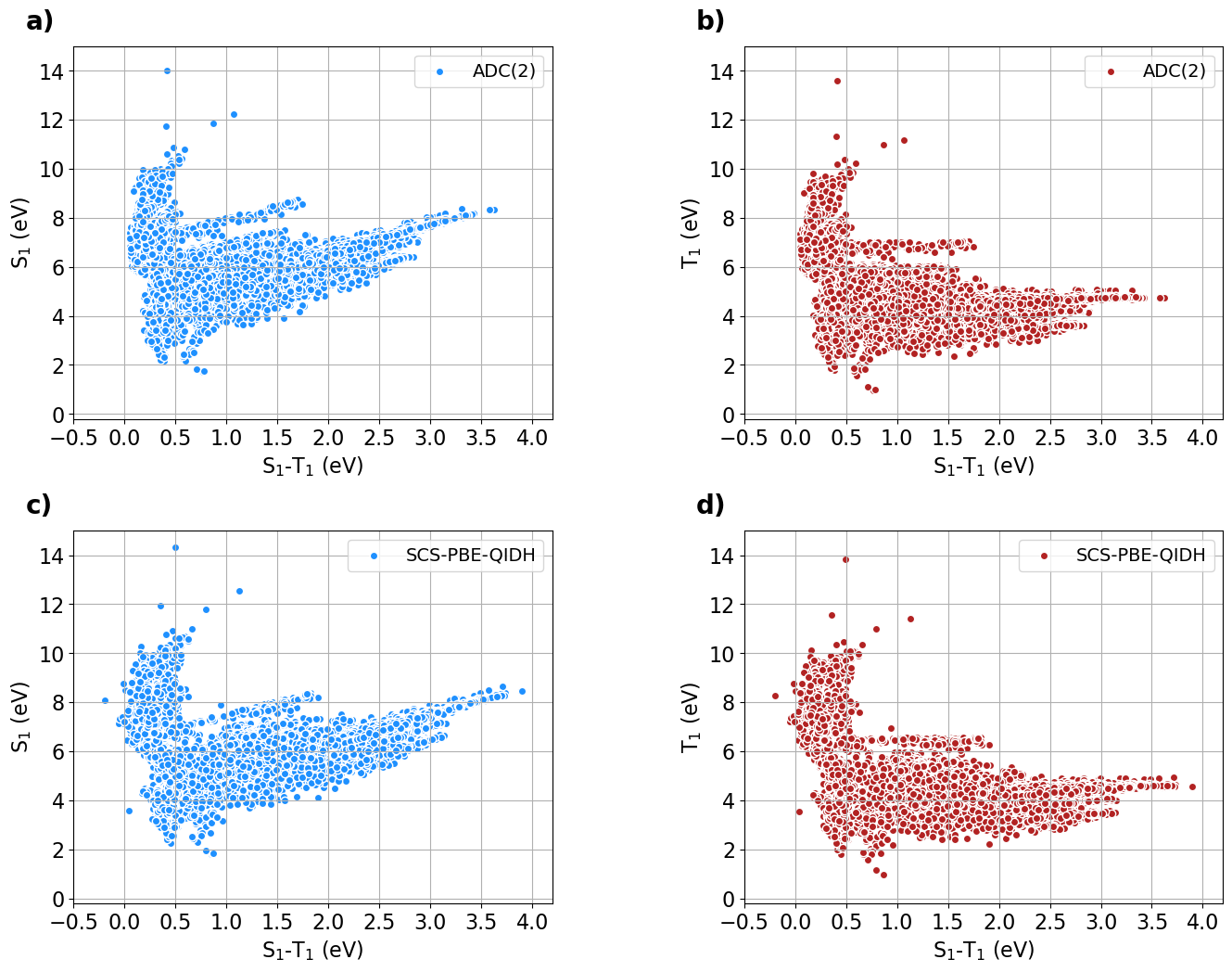}
    \caption{
    Distribution of S$_1$ and T$_1$ energies of 12,880 molecules 
    calculated with ADC(2) and SCS-PBE-QIDH methods
    shown jointly with the S$_1$-T$_1$ gap. 
    }
    \label{fig:SI_02}
\end{figure*}

\clearpage

% Fig. 3, query-1
\begin{figure*}[ht]
    \centering
    \includegraphics[width=\linewidth]{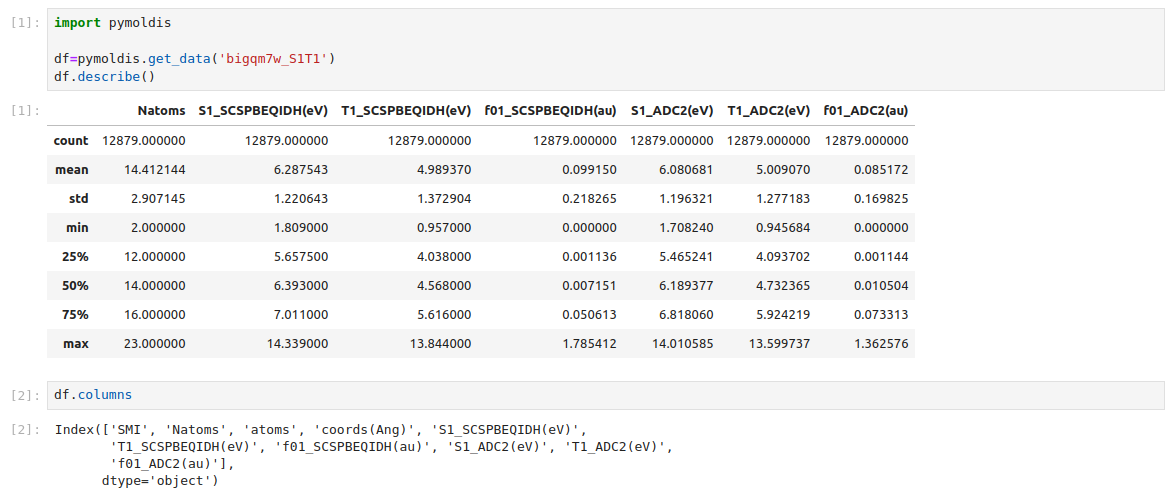}
    \caption{
    Example query 1: 
    Import the module {\tt pymoldis} in Python code, load the bigQM7$\omega$ dataset and perform a simple query 
    to get an overall summary of the dataset using {\tt .describe()} 
    and the names of all the columns using the
    {\tt columns} functionalities of {\tt Pandas} module. Screenshot of a Jupyter notebook available at
    {\tt https://github.com/moldis-group/pymoldis}. Note that, in {\tt pymoldis}, we have removed the entries 
    for the O$_2$ molecule as it is stable as a triplet in its ground state.
    }
    \label{fig:SI_03}
\end{figure*}

% Fig. 4, query-2
\begin{figure*}[ht]
    \centering
    \includegraphics[width=\linewidth]{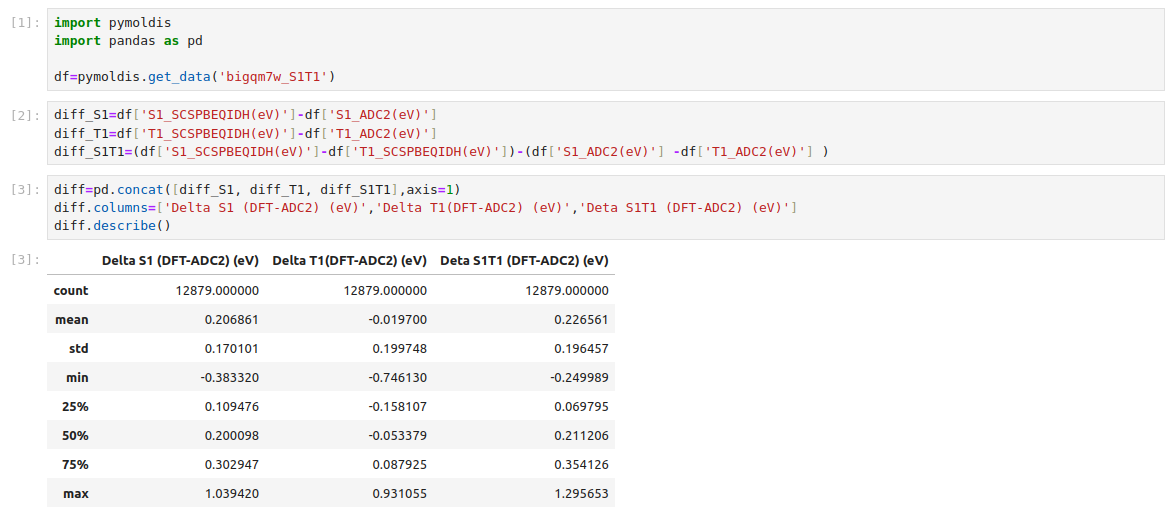}
    \caption{
    Example query 2: 
    Import {\tt pymoldis}, calculate the deviations of 
    SCS-PBE-QIDH predictions from ADC(2) values of 
    S$_1$, T$_1$, and S$_1$-T$_1$ energies, and get a summary of all three 
    deviations.
    }
    \label{fig:SI_04}
\end{figure*}

% Fig. 5, query-3
\begin{figure*}[ht]
    \centering
    \includegraphics[width=\linewidth]{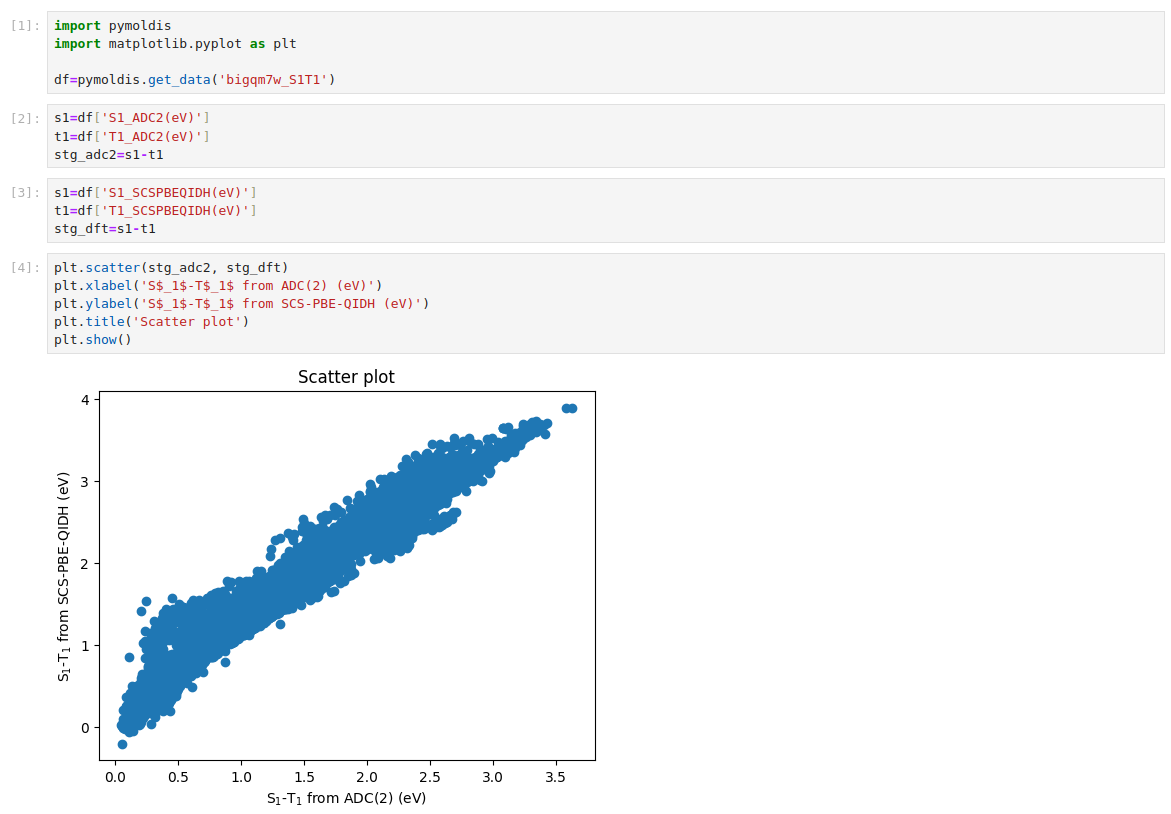}
    \caption{
    Example query 3:
     Import {\tt pymoldis}, load S$_1$-T$_1$ energies from the
     SCS-PBE-QIDH  and ADC(2) methods, and make a scatterplot.
    }
    \label{fig:SI_05}
\end{figure*}

% Fig. 6, query-4
\begin{figure*}[ht]
    \centering
    \includegraphics[width=\linewidth]{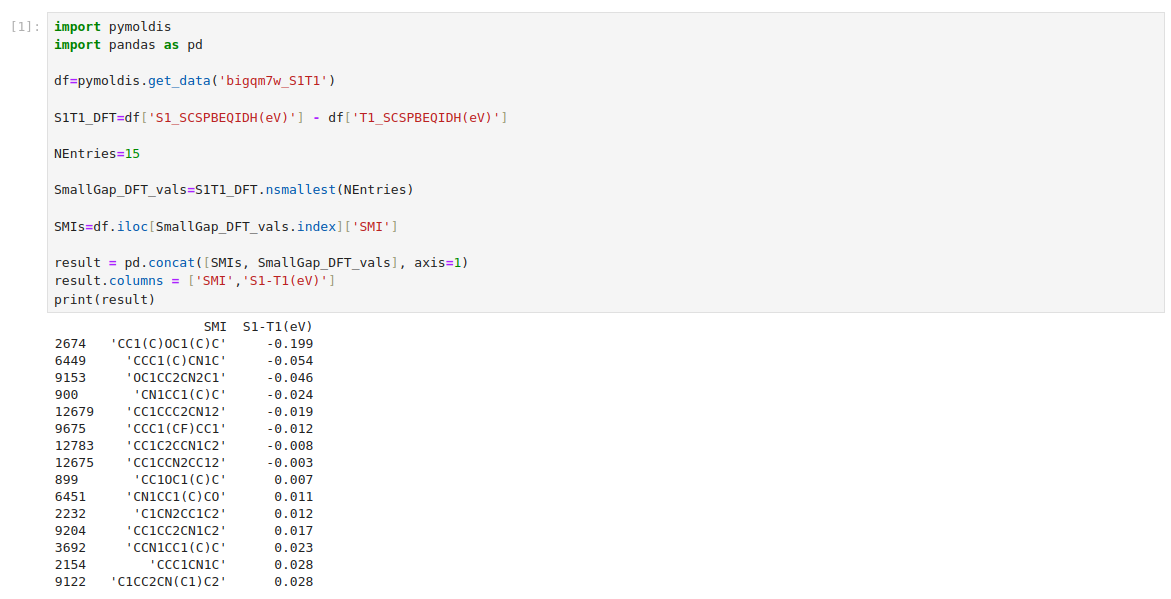}
    \caption{
     Example query 4: 
     Find 15 molecules in the bigQM7$\omega$ dataset with 
     the smallest S$_1$-T$_1$ energy gaps according to the
     SCS-PBE-QIDH/def2-TZVP method.
    }
    \label{fig:SI_06}
\end{figure*}

% Fig. 7, query-5
\begin{figure*}[ht]
    \centering
    \includegraphics[width=\linewidth]{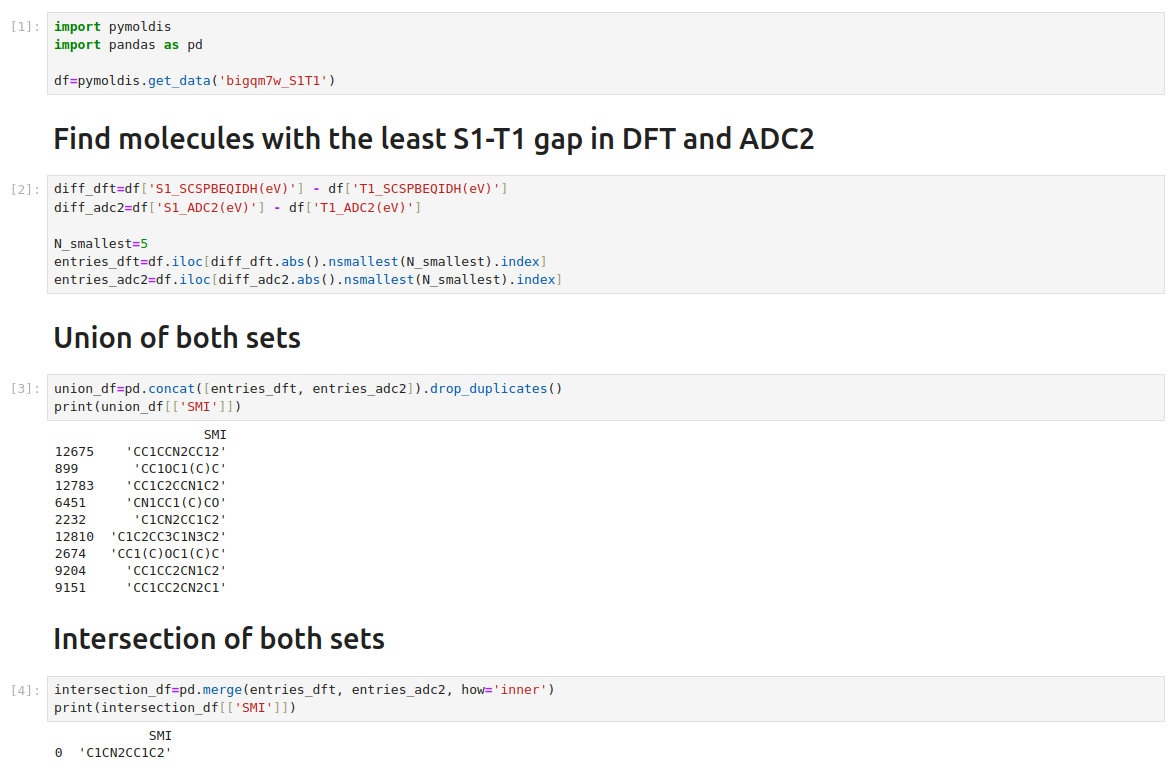}
    \caption{
    Example query 5:
    Find 5 molecules in the bigQM7$\omega$ dataset with 
     the smallest S$_1$-T$_1$ energy gaps according to the
     SCS-PBE-QIDH/def2-TZVP and the ADC(2) methods. Then,
     find the union and intersection of both sets. 
    }
    \label{fig:SI_07}
\end{figure*}

% Fig. 8, query-6
\begin{figure*}[ht]
    \centering
    \includegraphics[width=\linewidth]{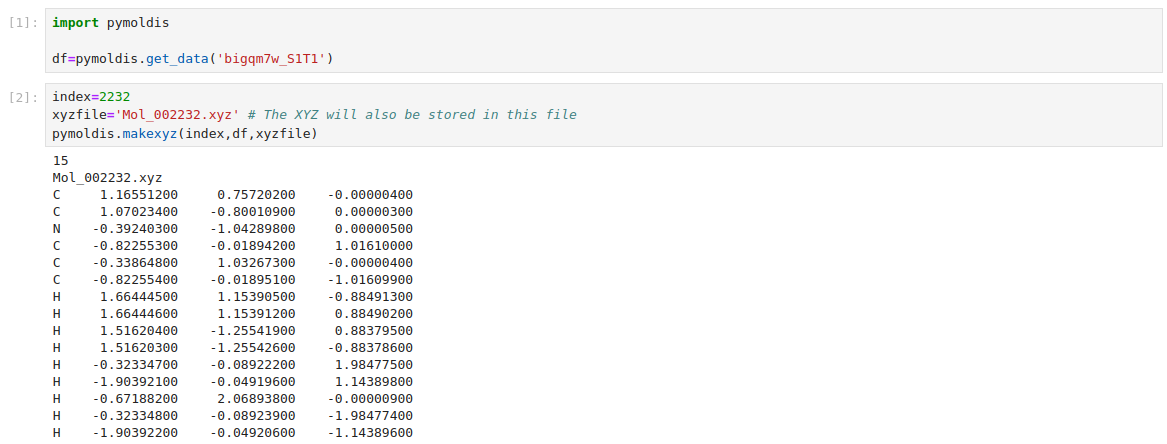}
    \caption{
    Example query 6:
    Get the Cartesian coordinates of the equilibrium geometry 
    of a molecule in the bigQM7$\omega$ dataset (determined at the $\omega$B97-XD/def2-TZVP level) using an index (perhaps one of them from the queries
    shown in Figure~S6 or Figure~S7.).
    }
    \label{fig:SI_08}
\end{figure*}

% Fig. 9, query-7
\begin{figure*}[ht]
    \centering
    \includegraphics[width=\linewidth]{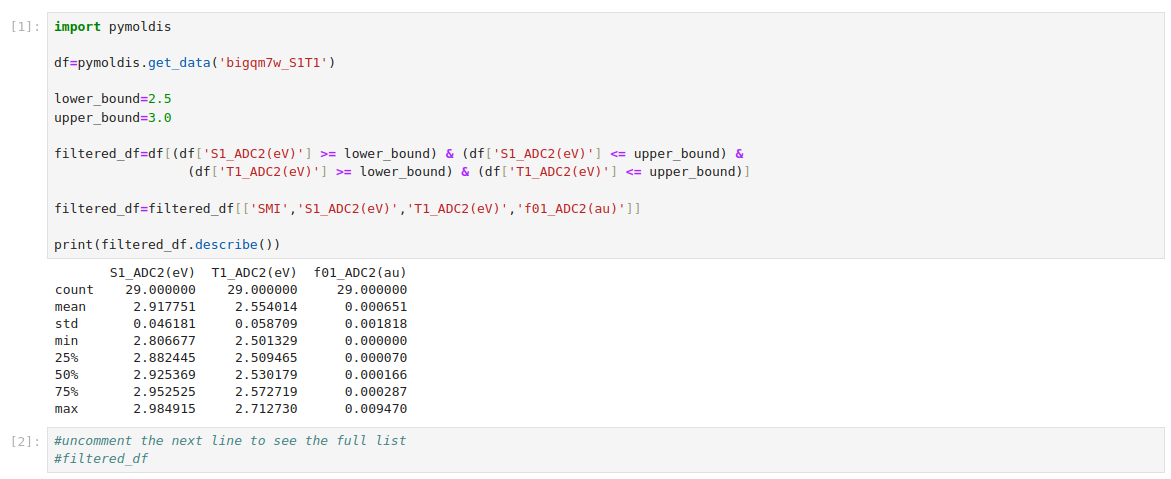}
    \caption{
    Example query 7: 
    Find molecules with S$_1$ and T$_1$ energies in the range
    2.5--3.0 eV, and print a summary.
    }
    \label{fig:SI_09}
\end{figure*}

% Fig. 10, query-8
\begin{figure*}[ht]
    \centering
    \includegraphics[width=\linewidth]{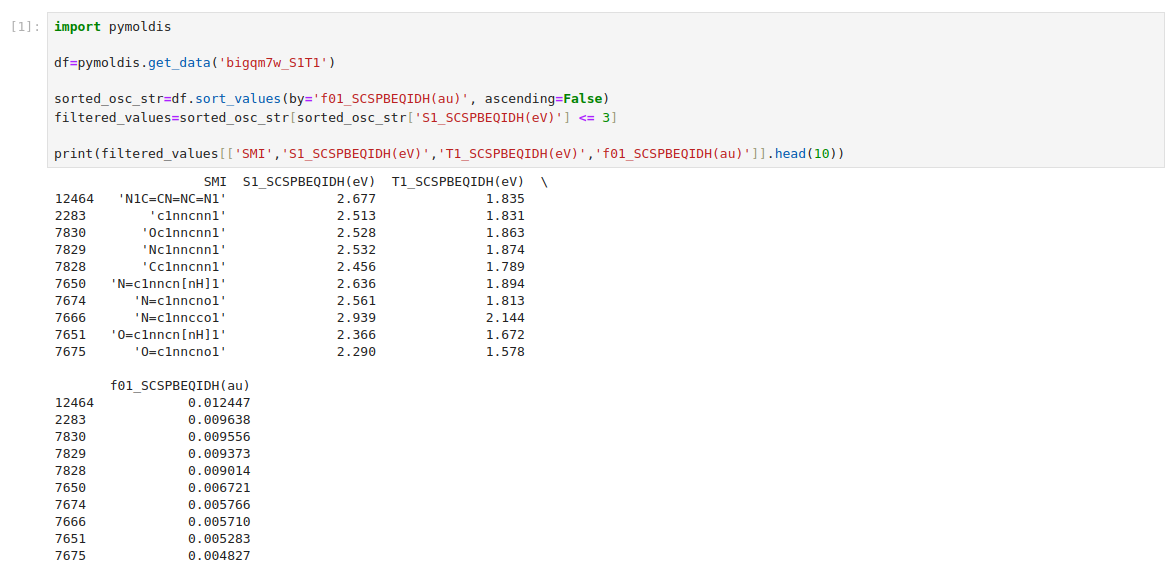}
    \caption{
    Example query 8:
    Find the bigQM7$\omega$ molecules with the largest oscillator strength
    for the S$_0$$\rightarrow$S$_1$ excitation, and print entries
    corresponding to the excitation energy of the S$_1$ state $\le$3 eV. 
    }
    \label{fig:SI_10}
\end{figure*}

% Fig. 11, query-9
\begin{figure*}[ht]
    \centering
    \includegraphics[width=\linewidth]{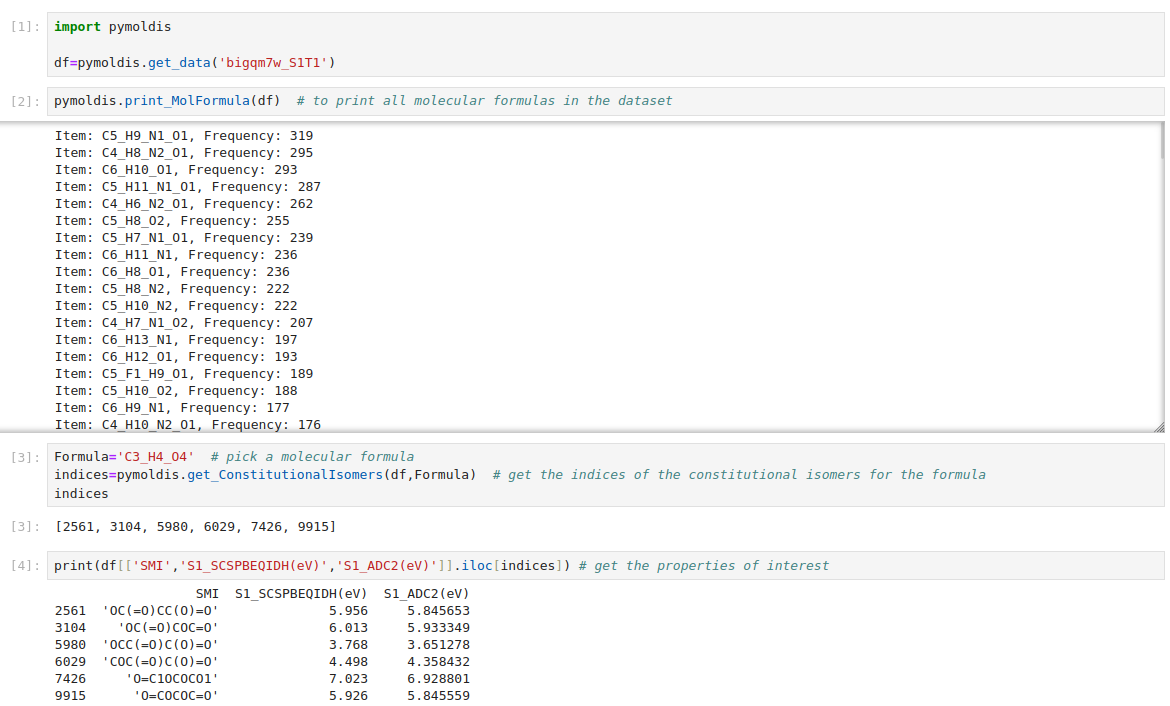}
    \caption{
    Example query 9: Get a list of all molecular formula (atomic compositions)
    spanned by the bigQM7$\omega$ molecules. Pick a molecular
    formula, and for the corresponding constitutional isomers,
    get indices, SMILES, and energetics.
    }
    \label{fig:SI_11}
\end{figure*}

\begin{figure*}[ht]
    \centering
    \includegraphics[width=\linewidth]{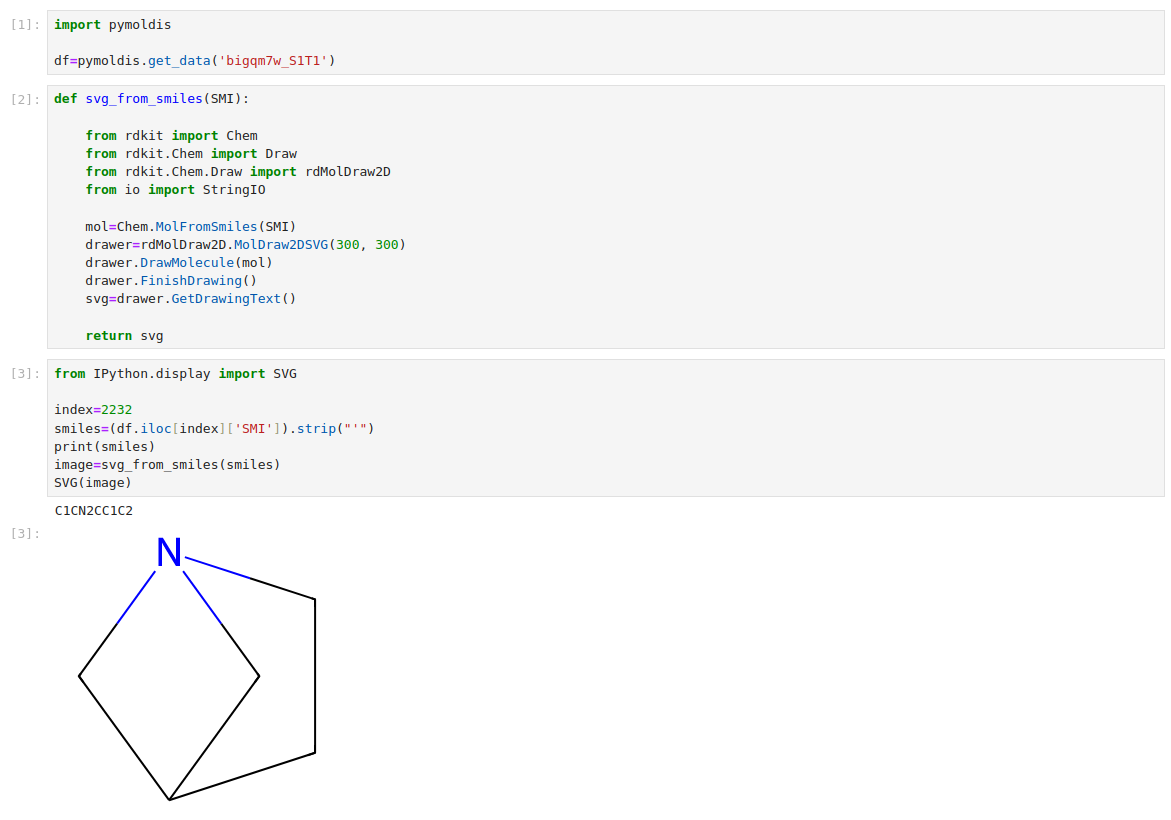}
    \caption{
    Example query 10: For an index in the bigQM7$\omega$ dataset,
    get the corresponding SMILES, and visualize the cartoon representation
    using the {\tt rdkit} module}.
    \label{fig:SI_12}
\end{figure*}

\clearpage

\singlespacing
\small
{ 
\begin{verbatim}
-----------------------------------------------------------------
MINIMUM ENERGY COORDINATES (ANGSTROEM) OF STRUCTURE 1 IN FIGURE 6
Calculated with wB97X-D3/def2-TZVP
-----------------------------------------------------------------
  N      0.390872    1.428585    1.624955
  C      0.708861    1.815707    0.253755
  C      0.575682    3.347565    0.045517
  C      0.089341    3.961412    1.361490
  H     -0.131796    3.573618   -0.756278
  H      1.533710    3.788806   -0.241382
  C      1.283999    2.122298    2.548251
  C      1.118564    3.662217    2.456078
  H      0.778668    4.076750    3.408734
  H      2.070037    4.145241    2.219926
  H      1.071210    1.764467    3.558078
  C     -0.992622    1.786176    1.923318
  C     -1.241399    3.307219    1.744630
  H     -0.039013    5.039944    1.249580
  H     -1.985937    3.494278    0.966417
  H     -1.626810    3.750993    2.666145
  H      1.723467    1.477122    0.032424
  H     -1.642047    1.200697    1.268645
  H     -1.207181    1.472821    2.947557
  H      2.307905    1.823158    2.312964
  H      0.038048    1.268226   -0.412274
-----------------------------------------------------------------
\end{verbatim}
}

\clearpage
\singlespacing
\small
{ 
\begin{verbatim}
-----------------------------------------------------------------
MINIMUM ENERGY COORDINATES (ANGSTROEM) OF STRUCTURE 2 IN FIGURE 6
Calculated with wB97X-D3/def2-TZVP
-----------------------------------------------------------------
  N      0.059194    1.259689   -0.034749
  C      0.181223    1.369404   -1.395803
  C     -0.296127    2.902657   -1.396472
  C      0.054220    3.314306    0.129364
  H     -1.363092    3.048798   -1.581413
  H      0.243182    3.522982   -2.117687
  C      1.181099    1.199343    0.750894
  C      1.520318    2.753136    0.525266
  H      1.915932    3.242972    1.419373
  H      2.223629    2.952868   -0.286825
  C      1.528474   -0.295853    0.278838
  H      0.899558    1.113411    1.801396
  C     -1.183797    1.210035    0.541175
  C     -1.055725    2.697881    1.134222
  H      0.050930    4.414046    0.217821
  H     -1.993011    3.260095    1.099742
  H     -0.688225    2.748816    2.162011
  C     -0.289605   -0.146435   -1.640642
  H      1.235079    1.394086   -1.677194
  C      0.064753   -0.795176   -0.200259
  H      0.247777   -0.642054   -2.453944
  H     -1.356951   -0.264126   -1.843230
  H      1.929624   -0.922619    1.080084
  H      2.228482   -0.357077   -0.558021
  C     -1.048484   -0.352316    0.888450
  H      0.066962   -1.894859   -0.288843
  H     -1.956302    1.270798   -0.226817
  H     -0.683096   -0.565357    1.895970
  H     -1.983093   -0.906070    0.762965
-----------------------------------------------------------------
\end{verbatim}
}

\clearpage
\singlespacing
\small
{ 
\begin{verbatim}
-----------------------------------------------------------------
MINIMUM ENERGY COORDINATES (ANGSTROEM) OF STRUCTURE 3 IN FIGURE 6
Calculated with wB97X-D3/def2-TZVP
-----------------------------------------------------------------
  N      0.379208    1.545025    1.615145
  C      0.522414    1.919464    0.194214
  C      0.040684    3.378560   -0.020385
  C      0.070623    4.079408    1.341912
  H     -0.971796    3.397890   -0.435621
  H      0.687101    3.882379   -0.742271
  C      1.397206    2.264418    2.406877
  C      1.412052    3.764525    2.012063
  H      1.569465    4.382515    2.898662
  H      2.237474    3.978652    1.325930
  H      1.089246    2.185002    3.450570
  C     -0.962585    1.942004    2.086010
  C     -1.045306    3.486323    2.208089
  H     -0.060595    5.156597    1.225223
  H     -2.028676    3.834363    1.884747
  H     -0.923759    3.801422    3.249156
  H      1.589067    1.872414   -0.031392
  H     -1.668148    1.617693    1.319548
  C     -1.351177    1.237061    3.350943
  C     -2.472023    0.550734    3.501386
  H     -0.672436    1.330637    4.194999
  H     -3.177772    0.431281    2.685513
  H     -2.727656    0.079541    4.442427
  C      2.748725    1.623174    2.311812
  C      3.451742    1.226245    3.359832
  H      3.172792    1.510693    1.317038
  H      4.436513    0.788344    3.252706
  H      3.063378    1.320067    4.368993
  C     -0.151387    0.943751   -0.723295
  C      0.442594    0.367902   -1.755661
  H     -1.200067    0.730826   -0.531445
  H     -0.090516   -0.308915   -2.411831
  H      1.487653    0.552956   -1.982883
-----------------------------------------------------------------
\end{verbatim}
}

\clearpage
\singlespacing
\small
{ 
\begin{verbatim}
-----------------------------------------------------------------
MINIMUM ENERGY COORDINATES (ANGSTROEM) OF STRUCTURE 4 IN FIGURE 6
Calculated with wB97X-D3/def2-TZVP
-----------------------------------------------------------------
  N      0.355841    1.689193    1.598935
  C      0.654217    1.567608    0.241850
  C      0.916390    3.184956    0.110907
  C      0.117101    3.739458    1.383964
  H      0.492086    3.569881   -0.818793
  H      1.953136    3.517290    0.137921
  C      1.382307    1.891585    2.521578
  C      0.949315    3.463932    2.724722
  H      0.351171    3.699694    3.603907
  H      1.834938    4.101303    2.773002
  C      1.206907    0.380433    3.106052
  C     -0.962678    1.564839    2.037086
  C     -1.333716    3.064287    1.474371
  H     -0.010959    4.829075    1.268782
  H     -1.838994    3.113681    0.510825
  H     -1.969501    3.596206    2.185549
  C      1.368005    0.121696    0.488440
  C      0.595592   -0.357605    1.814591
  H      2.440757    0.147130    0.671914
  H      1.195988   -0.565099   -0.340160
  H      0.525997    0.275976    3.949162
  H      2.159615   -0.058002    3.402458
  C     -0.972163   -0.025668    1.682064
  H      0.723680   -1.446292    1.930783
  H     -1.563157   -0.604598    2.391255
  H     -1.320345   -0.286401    0.684087
  C     -1.319803    1.664616    3.508357
  C     -2.446498    1.175892    4.005664
  H     -0.689579    2.244579    4.167756
  H     -3.139819    0.593203    3.409219
  H     -2.718670    1.354144    5.037962
  C      2.830237    1.995063    2.078788
  C      3.843275    1.830891    2.916769
  H      3.056792    2.300071    1.066856
  H      4.862687    2.000213    2.594920
  H      3.698317    1.535684    3.950444
  C     -0.433615    1.319396   -0.787093
  C     -0.169445    0.856936   -2.000303
  H     -1.452131    1.607214   -0.567215
  H     -0.950220    0.771167   -2.744907
  H      0.828234    0.554823   -2.300199
-----------------------------------------------------------------
\end{verbatim}
}

\onehalfspacing 
\clearpage 
\bibliography{ref}